\documentclass[fleqn,usenatbib]{mnras}

\usepackage{newtxtext,newtxmath}

\usepackage[T1]{fontenc}

\DeclareRobustCommand{\VAN}[3]{#2}
\let\VANthebibliography\thebibliography
\def\thebibliography{\DeclareRobustCommand{\VAN}[3]{##3}\VANthebibliography}

\usepackage{graphicx}	
\usepackage{amsmath}	
\usepackage{subcaption}
\usepackage{placeins}
\usepackage{float}
\usepackage{caption}

\title[Dark matter near Galactic Centre black hole]{Probing dark matter distributions with the pericentre precession of the stellar orbits near the Galactic Centre black hole}

\author[Paul and Kalita]{
Debojit Paul,$^{1}$\thanks{E-mail: debojitpaul645@gmail.com}
Sanjeev Kalita,$^{1}$
\\
$^{1}$Department of Physics, Gauhati University, Guwahati-781014, Assam, India
}

\date{Accepted XXX. Received YYY; in original form ZZZ}

\pubyear{\the\year{}}

\begin{document}
\label{firstpage}
\pagerange{\pageref{firstpage}--\pageref{lastpage}}
\maketitle

\begin{abstract}
The Galactic Centre black hole provides a naive environment for understanding unknown matter distribution and new gravitational physics. For this stellar orbits in the nuclear star cluster are reliable probes. We investigate different dark matter mass profiles through pericentre shift of stellar orbits near the black hole. We also study capability of existing and upcoming astrometric facilities to detect dark matter induced precession and to distinguish between several dark matter profiles. Parameters of different dark matter density profiles are estimated by using the most recent upper bound on dark mass near the black hole. These profiles are then used for calculating the gravitational potential and hence the relativistic pericentre shift of both low and high eccentricity orbits of 13 S-stars. We use the recently measured deviation parameter $f_{sp}$ for investigating competition between dark matter and gravitational physics within S2's orbit. The astrometric shift of the pericentres has been calculated and compared with existing and upcoming astrometric capabilities of large and extremely large telescopes. The orbit of S2 is found to be insensitive to dark matter induced precession. Low eccentricity and wider orbits are prominent probes for
measuring dark matter induced precession which is accessible to present and upcoming astrometric facilities such as Keck, GRAVITY and TMT. The existing and upcoming facilities can distinguish between different dark matter profiles for some stars and hence they posses the capability to distinguish between possible formation histories of the central region of our Galaxy.
\end{abstract}

\begin{keywords}
black hole physics -- gravitation -- dark matter -- Galaxy: centre --  stars: kinematics and dynamics
\end{keywords}



\section{Introduction}
Existence of supermassive black holes at the centres of galaxies was proposed by Lynden Bell and Rees \citep{Bell_Rees_1971}. This proposal founded the black hole paradigm of quasar activity. Early measurements of gas motions and radial velocity of stars at the centre of our Galaxy indicated existence of a dark central mass of $\sim 3 \times 10^6 \ M_\odot$ within a galactocentric radius of 0.1 pc \citep{Haller_1996,Genzel_1997}. This still allowed a central cluster of discrete objects with dark mass density of $10^9 \ M_\odot pc^{-3}$  with no assurance of presence of a black hole. However, advent of diffraction limited near infrared measurements indicated that a dark central mass of $(2.6 \pm 0.6) \times 10^6 M_\odot$  is confined within $0.015\ pc$. This corresponds to a dark density of $10^{12}M_\odot pc^{- 3}$ \citep{Eckart_Genzel_1997,Ghez_1998}. This ruled out the cluster hypothesis and was the first indication of existence of a gravitationally collapsed object, a supermassive black hole (will be mentioned henceforth as the Galactic Centre (GC) black hole) as proposed by Lynden Bell and Rees. 

Over the past three decades continuous high resolution observations of the Keck telescope and the Very Large Telescope (VLT) have revealed stellar motions in the innermost region of the Galactic Centre \citep{Eckart_Genzel_1997,Ghez_1998,Ghez_2003,Gravity_2018,Do_2019}. Orbits of more than 40 stars (known as S-stars) have been resolved \citep{Gillessen_2017,Peißker_2022,Gravity_2021,10.1093/mnras/stad3282}. Some of these stellar orbits have been used to probe the gravitational field of the GC black hole. Since 2017 the four telescope interferometric beam combiner GRAVITY/VLTI has achieved astrometric accuracy of $30-100\ \mu as$ and studied the orbits of the stars S2, S29, S38 and S55 \citep{GRAVITY2022}. Orbits of these stars are found to be consistent with the GC black hole mass $M = 4.30 \times 10^6\  M_\odot$. This is $0.25 \%$ precision measurement of the black hole mass. 

The GC black hole has been used as a naive laboratory to test gravity theory \citep{Hees_2017,Gravity_2020,ULDM2023,ULDM2023_PRD,10.1093/mnras/stad3282,GRAVITY2025}, spacetime metrics \citep{Zakharov_2018,Kalita_2023,Fernández_2023,Paul_2024,Moriyama_2025} and unknown mass distribution near the black hole \citep{GRAVITY2022,Heibel2022,GRAVITY2024,Lechien2024}. General relativistic effects have been tested by the GRAVITY Collaboration through measurement of gravitational redshift of light of the star S2 near its pericentre passage in 2018 and the detection of Schwarzschild in-plane precession of the star \citep{Gravity_2018,Gravity_2020}. These measurements were the first of their kinds to test General Relativity (GR) in a regime where gravitational field strength is some hundred order magnitude greater than the one we encounter in the solar system. It is not unreasonable to expect that the GC black hole environment can be probed by upcoming Extremely Large Telescopes (ELTs) to test some of the serious alternatives to GR. Theoretical calculations have been performed by several authors which demonstrate how the pericentre advance of compact stellar orbits can be used as a probe to constrain Yukawa gravitational potential of massive gravity theories and metric-$f(R)$ theories \citep{Hees_2017,Kalita2018,Kalita2020,Paul_2023,Paul_2024,GRAVITY2025}. Capability of stellar orbits to constrain unknown scalar fields near the black hole has also been explored \citep{Gravity_scalar_2019}. 

The Event Horizon Telescope (EHT) has produced the first ever images of the M87 black hole and the GC black hole \citep{EHT_2019,EHT_2022}. The angular size and shape of shadows cast by the two black holes have been measured. Black hole shadows are clean probes of horizon scale physics, particularly the test of black hole no-hair theorem and Kerr metric hypothesis. It is believed that all astrophysical black holes are described by the Kerr metric. The size of a shadow is $5GM/c^2$ for all spins \citep{Johannsen_2010}. Therefore, measurement of angular size of the black hole shadow is a null test of Kerr metric hypothesis \citep{Psaltis_2015}. According to the black hole no-hair theorem the exterior spacetime of a black hole is described only by mass and spin appearing in the Kerr metric. The consequence is that quadrupole moment ($q$) of a black hole is related to black hole spin ($\chi$) as $q = - \chi^2$.  Shape of a black hole shadow is circular for all but the highest spin. For any spacetime metric the shape depends on how much the quadrupole moment deviates from the Kerr value. A large asymmetry of the shadow indicates a violation of the no-hair theorem \citep{Psaltis_2016}.  As the mass of the GC black hole and its distance have been measured accurately in recent times, the precision in angular size of the shadow allows one to test strong field behavior of gravity and spacetime metric. 

In addition to presenting a laboratory for black hole metrics and gravitation theory, the GC black hole also gives us a new environment to understand mass distribution near the black hole contributed by stellar mass and unknown dark mass.  \cite{GRAVITY2022} produced the upper bound of $3000 M_\odot$ for this dark mass through the observation of S2’s orbit. In a recent study much more stringent bound has appeared as $1200 M_\odot$ \citep{GRAVITY2024}.

Cosmology offers a sound motivation for the study of dark matter near the GC black hole. Existence of supermassive black holes in the very early phase of the universe (at cosmological redshift $z >10$) has been confirmed by the James Webb Space Telescope (JWST). $34$ black holes of masses in the range $(10^{6.2\pm0.3}-10^{8.9\pm0.1}) M_\odot$ in the high $z$ universe are reported by JWST \citep{Dayal_2024}. This has confirmed that massive black holes must predate quasar activity and galaxy formation. Formation of supermassive black holes at the cosmic dawn constitutes an interesting open problem in cosmology.

One naive hypothesis is that a massive black hole can grow in the dark matter halo assembled by a Primordial Black Hole (PBH) seed. This is known as the “seed effect” \citep{Carr_Silk_2018}. The Carr-Silk mechanism advocates for massive PBH seed as generator of massive black holes and high redshift galaxies \citep{Liu_2022}. Another exotic process is the gravothermal collapse of Self Interacting Dark Matter (SIDM). SIDM can undergo gravothermal collapse leading to formation of massive black holes \citep{Ostriker_2000,Balberg_Shapiro_2002,Das_2024}. The SIDM possesses core like distribution. \cite{Das_2024} demonstrated possibility of growth of massive black hole seeds in the pre-quasar eras by incorporating gravothermal  collapse of SIDM with various density profiles and dark matter interaction cross sections guided by cosmology. Before these developments, \cite{Detweiler_1980} and \cite{Gondolo_silk} studied the interaction of massive black holes with dark matter. \cite{Gondolo_silk} proposed that if a supermassive black hole grows by adiabatic growth channel it can drag dark matter distribution onto it leading to a dark matter spike. The profile, known as Gondolo-Silk (GS) spike varies as $r^{-\gamma}$ with spike slope $2.25\le\gamma\le2.5$ \citep{Merritt_2002,10.1093/mnras/stad3282}. A spike profile can enhance the probability of dark matter annihilation if one assumes that the particle candidates of dark matter are supersymmetric such as neutralino \citep{Merritt_2004}. \cite{10.1093/mnras/stad3282} discussed the possibility of studying such a scenario with the help of available stellar orbits near the GC black hole.

\cite{Lacroix_and_silk_2013} predicted that a dark matter distribution near a supermassive black hole can affect the black hole shadow. Their model showed that EHT’s high resolution imaging of the GC black hole shadow can be utilized to constrain dark matter distribution. It has been shown by \cite{Xu_2018} that for a black hole embedded in a dark matter halo the exact form of space time metric can be predicted. This allows one to study effect of dark matter on black hole shadows \citep{Jusufi_2019}.

These cases are inspiring for investigation of effect of dark matter distribution on stellar orbits near the GC black hole. In-plane pericentre shift of stellar orbit acts as a probe for testing gravitation theory. The pericentre shift of S2 has been used to study compatibility of General Relativity near the GC black hole \citep{Gravity_2020}. It is also used as a probe for constraining a massive scalar mode in $f(R)$ gravity theory \citep{Kalita2020,Kalita_2021,Paul_2024}. In this work we investigate in-plane percentre shift of stellar orbits in presence of dark matter near the GC black hole. We compare the magnitude of retrograde shift of orbits due to dark matter with prograde shift due to general relativistic effect. We also compare the effect of dark matter with the effect of certain modifications to GR. We investigate the prospects of detection of these effects in the orbits of S-stars with existing and upcoming astrometric facilities. The paper is organized as follows. Section \ref{sec2} presents various dark matter mass profiles used in this work. In Section \ref{sec3} we present the pericentre shift induced by these dark matter profiles. The prospect of detection of these effects in the orbit of S-stars is highlighted in Section \ref{sec4}. Finally, we discuss the findings of this study and conclude in Section \ref{sec5}.

\section{Dark matter profiles}\label{sec2}
The standard Cold Dark Matter (CDM) paradigm in cosmology assumes that dark matter comprises some Weakly Interacting Massive Particles (WIMPs) \citep{Peebles_1982}. However, after several decades of experimental search existence of WIMPs has not yet been confirmed. Distribution of dark matter depends on what it is composed of. Therefore, there exists theoretical uncertainty on nature of the mass distribution of dark matter. Density profile of dark matter near the GC black hole depends on formation history of the inner region of our Galaxy \citep{Gnedin}. 
\par Presence of black hole at the Galactic Centre naturally modifies distribution of dark matter \citep{Detweiler_1980,Gondolo_silk}. Dark matter is likely to develop cusp or even spike with density profile which varies according to growth mechanism of the black hole and nature of the dark matter \citep{Gondolo_silk,Gnedin,Ullio_Zhao_2001}. It is also possible that dark matter particles follow an independent isotropic and quasi-equilibrium distribution due to dark matter interactions. In that case a black hole can be situated at the centre of the dark matter halo \citep{Alvarez_2021}. Not all the profiles considered in this work, therefore, are modified by the black hole. Probing unknown mass distribution near the supermassive black holes usually invokes two alternative density profiles. These are cored profiles (for example, Plummer type) which are used as benchmark for smooth finite distribution. Cuspy (Bahcall-Wolf) and Spike (Gondolo-Silk) distribution are result of modification due to gravity of the central black hole \citep{GRAVITY2022,GRAVITY2024}. Here we highlight some density profiles of interest.

\subsection{Plummer profile}
The Plummer profile was originally developed by \cite{plummer1911} for modeling globular cluster structures. It represents a spherically symmetric cored dark matter density distribution. It is given by,

\begin{equation}\label{eq1}
    \rho(r)=\rho_o \left(1+\frac{r^2}{r_o^2}\right)^{-\frac{5}{2}}
\end{equation}

where, $\rho_o$ and $r_o$ are density and scale parameter respectively. This analytical model has found renewed interest in modern astrophysics as a benchmark for testing extended mass distributions around supermassive black holes, particularly through precision observations of the S-stars \citep{GRAVITY2022,Heibel2022,Lechien2024,GRAVITY2024}. This profile is not modified by the black hole. Recent studies utilizing interferometric data from the GRAVITY instrument at the Very Large Telescope and Keck Observatory have tested the Plummer profile as one of the primary density models for constraining dark matter distributions within S2's orbit \citep{GRAVITY2022,GRAVITY2024}.

\subsection{Gnedin-Primack cusp profile}
The Gnedin \& Primack cusp (GPC) profile is a theoretical dark‑matter density profile that predicts a steep inner slope resulting from adiabatic contraction of the halo during baryonic infall \citep{Gnedin}. The density profile is given by,

\begin{equation}\label{eq1aa}
    \rho(r)=\rho_o \left(\frac{r}{r_{bh}}\right)^{-\frac{3}{2}} \ \ ; l<r\le r_{bh}
\end{equation}

Here, $l=10^{-5}$ pc and $r_{bh}$ is the sphere of influence of the central black hole. It describes a "mini-cusp" distribution that arises from two body scattering of dark-matter particles with the dense stellar cluster surrounding the GC black hole. This profile lies between the classical Navarro‑Frenk‑White (NFW) cusp ($\rho \propto r^{-1}$) and the even steeper spikes expected around supermassive black holes (see section \ref{GS_spike}). The central black hole's sphere of influence modifies the original halo configuration through stellar scattering. The nuclear star cluster, gas and stellar cusps constituting the complex baryonic environment near the GC black hole offers a natural setting to examine how baryonic processes might flatten or preserve the GP cusp.

\subsection{Bahcall-Wolf cusp profile}
The Bahcall-Wolf cusp (BWC) profile proposed by \cite{Bahcall_Wolf}, describes
the theoretically predicted stellar density distribution around a massive black hole that results from dynamical relaxation processes. This cusp is characterized by a power-law density profile. It is given by,

\begin{equation}\label{eq2}
    \rho(r)=\rho_o\left(\frac{r}{r_o}\right)^{-\frac{7}{4}}
\end{equation}

 The profile emerges when stellar orbits undergo energy exchange through two-body gravitational encounters over approximately one relaxation time, driving the system toward a
characteristic equilibrium state. N-body simulations have successfully
demonstrated the formation and growth of Bahcall-Wolf cusps in stellar systems taking into account close star-black hole interactions \citep{Preto_2004}. The model has gained recent attention in the context of Galactic Centre black hole as it can act as cold dark matter spike without particle self-annihilation \citep{Gondolo_silk,Gnedin,Sadeghian,Fields}, since its power-law slope falls comfortably within the observational limits of comparable power-law models.

\subsection{Gondolo-Silk (GS) spike profile}\label{GS_spike}
The inner dark‑matter distribution of galaxies remains one of the most uncertain aspects of cosmology, with large‑scale observations favoring flat (cored) profiles while numerical simulations predicting steep (cuspy) ones, a tension known as the cusp/core controversy \citep{de_Blok_2010}. The adiabatic growth of the GC black hole would compress the surrounding dark matter into a very steep “spike” \citep{Gondolo_silk}. The density profile for such spike is given by, \citep{Sadeghian}

\begin{equation}\label{eq3}
    \rho(r)=\rho_{halo}(R_{sp})\left(\frac{r}{R_{sp}}\right)^{-\gamma_{sp}} \ \ ;\ 2 R_{Sch} \leq r < R_{sp}
\end{equation}

where, $R_{Sch}=2GM_{BH}/c^2$ is the Schwarzschild radius and $R_{sp}$ is the radial extension of the spike. $\rho_{halo}(R_{sp})$ is the halo density and spike radius ($R_{sp}$) and is defined by a generalized Navarro-Frenk-White (NFW) profile. It is characterized by the parameter $\gamma$ and is given by,

\begin{equation}\label{eq4}
    \rho_{halo}(r)=\rho_s \left(\frac{r}{r_s}\right)^{-\gamma}\left(1+\frac{r}{r_s}\right)^{\gamma-3}
\end{equation}

where, $r_s$ is the scale radius and the scale density $\rho_s$ is related to background density $\rho_\odot$ as

\begin{equation}\label{eq5}
    \rho_s=\rho_\odot\left(\frac{R_o}{r_s}\right)^\gamma\left(1+\frac{R_o}{r_s}\right)^{3-\gamma}
\end{equation}

Spike density is usually thought to facilitate dark matter annihilation \citep{Ullio_Zhao_2001,Gnedin,Dosopoulou_2025}. The spike can dramatically enhance annihilation signals in the region offering a unique laboratory for indirect dark‑matter searches. However, dynamical processes such as mergers, non‑adiabatic black‑hole growth, or heating by the stellar cusp can weaken or erase the spike \citep{Merritt_2002}. The observed lack of a stellar spike near the GC black hole does not automatically exclude a dark‑matter spike because the two components can decouple \citep{Lacroix_2018}. Consequently, constraining the spike profile with S-star orbits offer a rare, model‑independent way to address the cusp/core problem and to assess the viability of dark‑matter annihilation scenarios near the GC black hole \citep{Lacroix_2018,10.1093/mnras/stad3282}.

\subsection{Self-interacting Dark Matter profile}
Self-interacting dark matter (SIDM) proposes that dark-matter particles scatter elastically with a cross-section per unit mass of approximately $0.1 - 10\ cm^2\ g^{-1}$ \citep{Rocha_2013,Harvey_2015,Loeb_2022}. This process can thermalise halo centres, transforming steep cusps in collisionless cold-dark-matter simulations into constant-density cores. The size of the core scales with cross-section of self interaction. The density is sensitive to heat conduction driven by self‑scattering, which can erase or soften spikes over a relaxation time comparable to the age of the galaxy \citep{Alvarez_2021}.

\cite{Pollack_2015} suggested that SIDM can form seed black holes in the centre of a halo by the process of gravothermal collapse. For reasonable value of self-interaction cross section and the dark matter mass fraction of the SIDM component these seed black holes can grow upto $10^9\ M_\odot$ black holes by cosmological redshift $z\ge 6$. This explains existence of high redshift massive black holes which is otherwise a tension for the standard $\Lambda$CDM cosmology.

\cite{Feng_2021} showed that gravothermal evolution of SIDM joined by baryons can accelerate the growth of massive black holes at $z=6-7$. Therefore, if the GC black hole evolved via process involving SIDM the orbits of S-stars may provide clue to constraints on such a candidate of dark matter. One interesting aspect of SIDM is that indirect‑detection signals ($\gamma$‑rays, neutrinos) from annihilation are strongly attenuated in SIDM cores. Thus, the S-star orbits provide a unique opportunity to study these properties in the vicinity of the GC black hole. The SIDM halo undergoes gravothermal collapse and produces isothermal cores for moderate cross sections. For higher cross sections however, the SIDM halo collapses (gravothermally and not due to the central black hole) into a steep cusp \citep{Dutra_2025}. Hence, the central black hole does not play any role in modifying the profile. In this work we consider two popular SIDM profiles which are presented below.

\subsubsection{$r^2$ model}
 The core of an SIDM halo is usually defined by a modified-core isothermal profile. At the inner core it behaves as an isothermal spherical profile with a constant density \citep{Begeman_1991}. The density profile is given by,

\begin{equation}\label{eq6}
    \rho(r)=\frac{\rho_o r_o^2}{r^2 + r_o^2}
\end{equation}

\subsubsection{$r^3$ model}
Another modified-core isothermal profile was proposed by \cite{Brownstein_2009}.
The density profile is given by,

\begin{equation}\label{eq7}
    \rho(r)=\frac{\rho_o r_o^3}{r^3 + r_o^3}
\end{equation}

\subsection{Fuzzy Dark Matter profile}\label{sec2.5}
Fuzzy dark matter (FDM) is a class of ultralight bosonic dark‑matter candidate with particle masses $\lesssim 10^{-22}$ eV, for which the associated de Broglie wavelength is large ($\sim$ kpc), causing the dark matter to behave as a coherent quantum wave rather than a collection of collisionless particles \citep{Hu_Barkana_Gruzinov_2000,Hui_2017}. The wave nature introduces an effective quantum‑pressure term that smooths density fluctuations below the de Broglie scale, leading to a suppression of the linear matter power spectrum on small scales and the formation of solitonic cores at the centre of halos \citep{Schive_2014,Simon_Springel_2021}. Cosmological simulations have demonstrated a cutoff in the halo‑mass function and a reduced abundance of low‑mass halos, which can alleviate the missing‑satellite, cusp‑core, and too‑big‑to‑fail problems of the standard $\Lambda$CDM model \citep{Schive_2014,Simon_Springel_2021}. At larger scales the model reproduces the successful $\Lambda$CDM predictions for the cosmic‑microwave‑background anisotropies and large‑scale structure. Lyman‑$\alpha$ forest measurements and ultraviolet luminosity function (ULVF) have constrained boson mass to be $\gtrsim 10^{-22}$ eV when quantum‑pressure effects are included \citep{Zhang_2018,Sipple_2025}. Analytically it has been demonstrated that “gravitational atom”(bound states of scalar particles) can form near the horizon due to presence of central black hole which contributes a very small potential term \citep{Bucciotti_2023}. But it has been seen that the density spike near the horizon is negligible compared to the solitonic core once the self-gravity of the soliton is included \citep{Alcubierre_2025}. Hence, the central black hole can add a very small perturbation. However, it cannot reshape the solitonic core into a new profile. The stellar orbits near Sgr A* present an unique opportunity to investigate predictions of FDM \citep{ULDM2023,ULDM2023_PRD,Chan_Lee_2022_ULDM}. An FDM core would affect the stellar trajectory which can be detected observationally. The density profile of the solitonic cores can be approximated as \citep{Schive_2014,Mocz_2017}

\begin{equation}\label{eq8}
    \rho(r)=\frac{\rho_o}{(1+Ar^2)^8}
\end{equation}

\noindent where, $\rho_o$ is the core density of the halo given by,

\begin{equation}\label{eq9}
    \rho_o=1.9\left(\frac{M_a}{10^{-23} \text{eV}}\right)^{-2}\left(\frac{r_c}{\text{kpc}}\right)^{-4} \frac{M_\odot}{\text{pc}^3}
\end{equation}

where, $M_a$ is the boson mass and $r_c$ is the core radius which is related to the parameter $A$ as $A=9.1\times 10^{-2}/r_c^{2}$. Also, the core radius depends on halo mass, $M_{halo}$ given by, \citep{Schive_2014b}

\begin{equation}\label{eq10}
    r_c=1.6\left(\frac{M_a}{10^{-22} \text{eV}}\right)^{-1}\left(\frac{M_{halo}}{10^9 M_\odot}\right)^{-1/3}\ \text{kpc}
\end{equation}

\section{Methods and calculations: Pericentre shift of stellar orbits due to dark matter}\label{sec3}

The method of obtaining the effect of dark matter distribution is as follows: 
We take relativistic pericentre shift of stellar orbits as a probe. For calculating pericentre shift due to extended mass distribution which is present near the GC black hole, we adopt the method given by \cite{PhysRevD.75.082001}. The pericentre shift depends on the gravitational potential of the mass distribution which in turn is governed by dark matter density profiles. Parameters of the density profiles are fixed by the recently measured \citep{GRAVITY2024} upper bound on total dark mass that is allowed to exist within the apocentre of the S2 star. We take orbital data for 13 S-stars (see Table \ref{tabA}) \citep{Gillessen_2017,Peißker_2020a,Peißker_2020b,Peißker_2022,GRAVITY2024}. Few of them have already crossed their pericentres. Others are going to have pericentre passage within the coming decade. The calculated pericentre shift is compared with the Schwarzschild precession of the orbits of these stars. We try to identify stellar orbits which are sensitive to effect of dark matter. The astrometric shift of the pericentre due to general relativistic and dark matter effect is also calculated and compared with capability of Keck, GRAVITY instrument in the Very Large Telescope (VLT) and the upcoming Thirty Meter Telescope (TMT). The mathematical outline of the calculations is shown below:

For any density distribution $\rho(r)$, the extended mass distribution is estimated as,

\begin{equation}\label{eq12}
    M(r)=\int_{r_{min}}^{r_{max}}4\pi r'^2\rho(r')dr'
\end{equation}

This extended mass distribution produces a gravitational potential which has the form,

\begin{equation}\label{eq13}
    \Phi_{ext}(r)=-G\int_{r}^{\infty}\frac{M(r')}{r'^2}dr'
\end{equation}

\cite{PhysRevD.75.082001} developed a method where the secular precession due to any central force (Newtonian) perturbation can be calculated. Any extended mass distribution near the GC black hole can be treated as a small classical perturbation to the Keplerian orbits of S-stars (see for example, \cite{Jovanovic_et_al_2021,Heibel2022,Tomaselli_2025} ). We identify that this potential is a small perturbation in addition to the $1/r$ Newtonian term, the total potential ($\Phi_{total}(r)$) of the system is expressed as,

\begin{equation}\label{eq14}
    \Phi_{total}(r)=\Phi_N(r) + \Phi_{ext}(r)
\end{equation}

where, $\Phi_N(r)=-GM/r$ is the Newtonian potential. This gives the perturbed potential as,

\begin{equation}\label{eq15}
    V(r)=\Phi_{total}(r)-\Phi_N(r)=\Phi_{ext}(r)
\end{equation}

The rate of precession caused by this perturbed potential can be estimated as,

\begin{equation}\label{eq16}
    \Delta\theta=-\frac{2 L}{G M_{bh} e^2}\int_{-1}^{1}\frac{zdz}{\sqrt{1-z^2}}\frac{dV(z)}{dz}
\end{equation}

where, $r$ is related to $z$ as $r=\frac{L}{1+e z}$. Here, $L=a(1-e^2)$ is the semi-latus rectum of the orbit having semi-major axis $a$ and eccentricity $e$. $M_{bh}$ is the mass of the central black hole.

\subsection{Constraints on parameters of the model}
Any extended mass distribution within the apocentre of S2 ($1942.86$ au) was constrained to an upper limit of $3000\ M_\odot$ ($0.1\%$ of $M_{bh}$) with $1\sigma$ confidence \citep{GRAVITY2022}. This upper limit was further reduced to $1200\ M_\odot$ ($0.028\%$ of $M_{bh}$) with $1\sigma$ confidence \citep{GRAVITY2024}. From this analysis it was concluded that this mass bound is the most stringent upper limit for any extended mass distribution. It is evident from the mass profiles displayed above that the parameters of the model govern the total mass budget within a given scale. Therefore, the model parameters are constrained such that the total mass within the apocentre of S2 is within $1200 M_\odot$. This can be written as,

\begin{equation}\label{eq20}
    M(r_a)=1200\ M_\odot
\end{equation}

\noindent Where, $r_a=1942.86$ au is the apoapsis distance of S2. The parameter $r_o$ appearing in the Plummer, Bahcall-Wolf and SIDM profile is chosen to be $0.3'' \approx 2475.18$ au which is of the order of $r_a$ \citep{GRAVITY2022,Heibel2022,GRAVITY2024}. Using equation (\ref{eq20}), the model parameters are calculated as follows:
\subsubsection*{Plummer profile:}
For the Plummer profile using equation ($\ref{eq1}$), the mass distribution is worked out as:

\begin{equation}\label{eq21}
    M(r)=\int_0^{r}4 \pi r'^2\rho(r')dr'=\frac{4}{3}\pi \rho_o\frac{r_o^3 r^3}{(r^2+r_o^2)^{3/2}}
\end{equation}

Using equation (\ref{eq20}) and substituting the values of $r_a$ and $r_o$, $\rho_o$ is estimated to be $4.786 \times 10^{-11}$ $kg/m^3$.

\subsubsection*{Gnedin-Primack cusp profile:}
The S-star - Sgr A* system fall within the sphere of influence of GC black hole ($r_{bh}$). Hence, the mass profile from equation (\ref{eq1aa}) is worked out as:

\begin{equation}\label{eq1aa_m}
    M(r)=\int_{l}^{r}4\pi r'^2 \rho(r')dr'=\frac{8}{3}\pi \rho_o r_{bh}^{3/2}(r^{3/2}-l^{3/2})
\end{equation}

The sphere of influence ($r_{bh}$) is $2$ pc \citep{Gnedin}. Substituting the value of $r_a$, $r_{bh}$ and $l$ in equation (\ref{eq1aa_m}), $\rho_o$ is estimated as $\rho_o=3.778\times 10^{-15}\ kg/m^3$.

\subsubsection*{Bahcall-Wolf cusp profile:}
The Bahcall-Wolf cusp mass profile is worked out using equation (\ref{eq2}). It has the form:

\begin{equation}\label{eq22}
    M(r)=\int_{0}^r 4\pi r'^2 \rho(r')dr'=\frac{16}{5}\rho_o r_o^{7/4}r^{5/4}
\end{equation}

Substituting $r_a=1942.86$ au and $r_o=2475.18$ au in equation (\ref{eq20}),  $\rho_o$ is estimated as $6.341\times10^{-12}$ $kg/m^3$.

\subsubsection*{Gondolo-Silk spike profile:}
The mass profile for Gondolo-Silk spike is estimated from equation (\ref{eq3}). It has the form:

\begin{multline}\label{eq23}
    M(r)=\int_{2R_{sch}}^{r}4\pi r'^2\rho(r')dr'\\=\frac{4\pi \rho_{halo}(R_{sp}) R_{sp}^{\gamma_{sp}}}{3-\gamma_{sp}}\left[r^{3-\gamma_{sp}}-(2R_{sch})^{3-\gamma_{sp}}\right]
\end{multline}

Using the VLT and KECK data for S-stars, \cite{10.1093/mnras/stad3282} constrained the parameters $R_{sp}\approx 15.7$ pc and $\gamma_{sp} \approx 2.32$. Therefore, using equation (\ref{eq20}), $\rho_{halo}(R_{sp})$ is estimated as $1.762\times10^{-19}$ $kg/m^3$.

\subsubsection*{SIDM ($r^2$) profile:}
The mass profile for SIDM ($r^2$) profile is estimated using equation (\ref{eq6}). It is expressed as:

\begin{equation}\label{eq24}
    M(r)=\int_{0}^{r}4\pi r'^2 \rho(r')dr'=4\pi \rho_o r_o^2\left(r-r_o \tan^{-1}\left(\frac{r}{r_o}\right)\right)
\end{equation}

Substituting $r_a=1942.86$ au and $r_o=2475.18$ au in equation (\ref{eq20}),  $\rho_o$ is estimated to be $3.151\times10^{-11}$ $kg/m^3$.

\subsubsection*{SIDM ($r^3$) profile:}
The mass profile for SIDM ($r^3$) profile is estimated using equation (\ref{eq7}). It is expressed as:

\begin{equation}\label{eq25}
    M(r)=\int_{0}^{r}4\pi r'^2 \rho(r')dr'=\frac{4}{3}\pi \rho_o r_o^3 \log\left(1+\frac{r^3}{r_o^3}\right)
\end{equation}

Substituting $r_a=1942.86$ au and $r_o=2475.18$ au in equation (\ref{eq20}),  $\rho_o$ is estimated to be $2.86\times10^{-11}$ $kg/m^3$.

\subsubsection*{FDM profile:}
For the FDM profile (\ref{eq8}), the mass profile is worked out as:

\begin{multline}\label{eq26}
    M(r)=\int_{0}^{r}4\pi r'^2 \rho(r')dr'=\frac{4\pi\rho_o}{215040 A^{3/2}}
    (\sqrt{A}r(
    3465A^6r^{12}\\
    +23100A^5r^{10}
    +65373A^4r^8
    +101376A^3r^6
    +92323A^2r^4\\
    +48580Ar^2
    -3465)/(1+Ar^2)^7
    +3465 \tan^{-1}(\sqrt{A}r))    
\end{multline}

The parameters $\rho_o$ and $A$ are both connected to the boson mass $M_a$ (see section \ref{sec2.5}). The Milky Way halo mass ($M_{halo}$) is taken to be $1.08\times10^{12} M_\odot$ from $Gaia$ DR2 \citep{10.1093/mnras/staa1017}. The parameters of the model can be expressed as,

\begin{equation}\label{eq27}
    \rho_o=2.164\times10^{-20}\ M^2\ \ kg/m^3\  ;\ A=6.127\times10^{-41}\ M\ \ m^{-2}
\end{equation}

where, $M=M_a/10^{-23} eV$ is a dimensionless quantity. The equation (\ref{eq26}) becomes a transcendental equation in $M$. This is solved numerically and the value of $M$ is found to be $M=32870.7$. This gives the boson mass as,

\begin{equation}\label{eq28}
    M_a=3.287\times 10^{-19}\ \text{eV}
\end{equation}

This result is consistent with the bound on ultra light bosons reported by \cite{ULDM2023_PRD} and \cite{ULDM2023}. The parameters then take the values,

\begin{equation}\label{eq29}
     \rho_o=2.338\times10^{-11} \ kg/m^3\  ;\ A=2.014\times10^{-36} \ m^{-2}
\end{equation}

\begin{table}
\centering
\caption{Model parameters for dark matter density profiles.}
\label{tab1}
\resizebox{0.4\textwidth}{!}{%
\begin{tabular}{ccc}
\hline
\multicolumn{2}{c}{Model}          & Parameters \\ \hline
\multicolumn{2}{c}{Plummer}        & \begin{tabular}[c]{@{}c@{}}$r_o=2475.18$ au $= 3.7\times10^{14}\ m$,\\ $\rho_o=1.69\times10^{-10}$ $kg/m^3$\end{tabular}  \\ 
\multicolumn{2}{c}{Gnedin-Primack cusp}  & \begin{tabular}[c]{@{}c@{}}\\$r_{bh}=2$ pc $= 6.15\times10^{16}\ m$ ,\\ $l=10^{-5}$ pc $=3.07\times 10^{11}\ m$ , \\ $\rho_o=1.69\times10^{-10}$ $kg/m^3$\end{tabular}  \\ 
\multicolumn{2}{c}{Bahcall-Wolf}   & \begin{tabular}[c]{@{}c@{}}\\ $r_o=2475.18$ au $= 3.7\times10^{14}\ m$,\\ $\rho_o=6.341\times10^{-12}$ $kg/m^3$\end{tabular} \\ 
\multicolumn{2}{c}{Gondolo - Silk} &
  \begin{tabular}[c]{@{}c@{}}\\ $\gamma_{sp}=2.32$,\\ $R_{sp}=15.7$ pc$=4.8\times 10^{17}\ m$,\\ $\rho_{halo}(R_{sp})=1.762\times10^{-19}$ $kg/m^3$\end{tabular} \\ 
\multicolumn{2}{c}{SIDM ($r^2$)} &
  \begin{tabular}[c]{@{}c@{}}\\ $r_o=2475.18$ au $= 3.7\times10^{14}\ m$ ,\\ $\rho_o=3.151\times10^{-11} kg/m^3$\end{tabular} \\ 
\multicolumn{2}{c}{SIDM ($r^3$)} & \begin{tabular}[c]{@{}c@{}}\\ $r_o=2475.18$ au $= 3.7\times10^{14}\ m$,\\ $\rho_o=2.86\times10^{-11}$ $kg/m^3$\end{tabular}  \\
\multicolumn{2}{c}{FDM} &
  \begin{tabular}[c]{@{}c@{}}\\$M_a=3.287\times10^{-19}$ eV,\\ $A=2.014\times10^{-36}$ $m^{-2}$,\\ $\rho_o=2.338\times10^{-11}$ $kg/m^3$\end{tabular} \\ \hline
\end{tabular}%
}
\end{table}

\begin{figure*}
    \centering
    \begin{subfigure}[b]{0.49\textwidth}
        \centering
        \includegraphics[width=\textwidth]{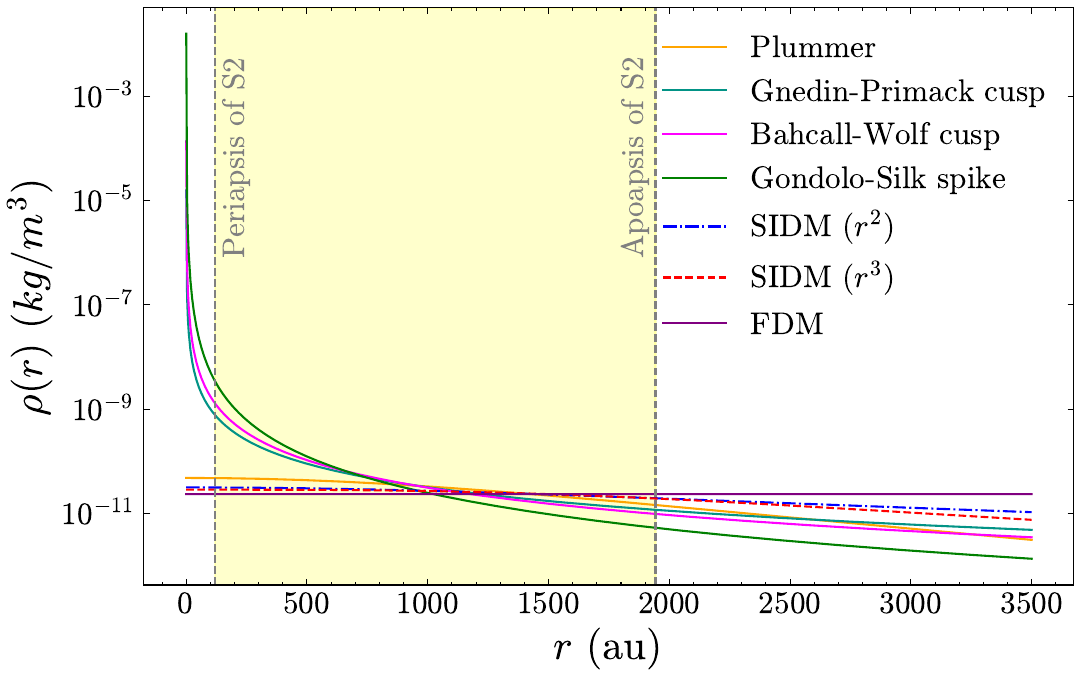}
        \caption{}
        \label{fig:1}
    \end{subfigure}
    \begin{subfigure}[b]{0.49\textwidth}
        \centering
        \includegraphics[width=\textwidth]{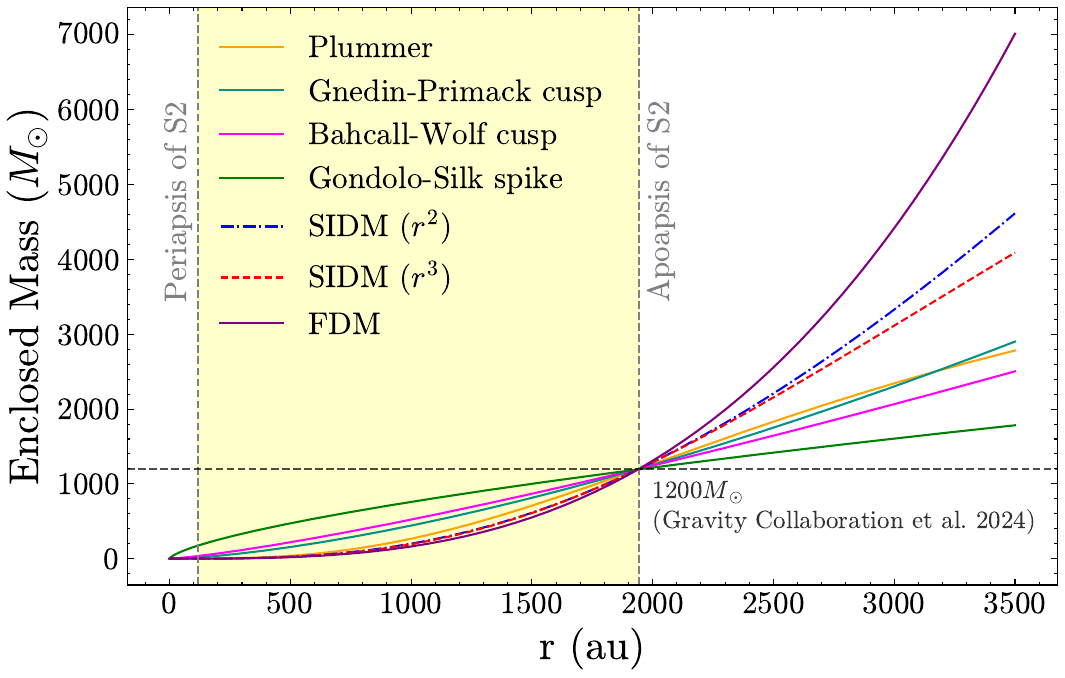}
        \caption{}
        \label{fig:2}
    \end{subfigure}
\caption{Variation of density profiles (panel (a)) and mass profiles (panel (b)) with galactocentric distance. The grey vertical lines represent pericentre and apocentre of S2. The black horizontal line in panel (b) represents the $1200\ M_\odot$ bound.}
\label{fig1aa}
\end{figure*}

The parameters for the different models considered in this study are summarized in Table \ref{tab1}. With these parameters the density and mass profiles are plotted in Figure \ref{fig1aa}.

It is important to note that the parameters of dark matter mass profiles are not taken from rotation curve (RC) data. It is challenging to determine RC near the Galactic Centre black hole. \cite{Sofue_2013} generated a grand rotation curve by combining disk to halo RC, bulge to disk RC and measuring RC in the region $(1-10)$ pc near the black hole. Information could only be obtained for de-Vaucouleurs profile and exponential bulge profile. These profiles are not of interest for our study. We are interested to study profiles which are closely related to presence and formation history of the black hole. Several profiles are not at all associated with flat RC as they are modified by the black hole. They are rather interesting for other aspects such as constraining annihilation of WIMP like dark matter \citep{Gondolo_silk}. Cuspy and spike profiles are of this category.

SIDM does not destroy success of CDM. The SIDM distribution in the large scale is given by Navarro-Frenk-White (NFW) profile \citep{Rocha_2013}. SIDM reduces to CDM cusp predicted by NFW profile in the inner region of a halo. This indeed explains flatter core demanded by galaxy RC \citep{Salucci_2000,Salucci_2012}. The soliton core of FDM is enveloped by an NFW profile \citep{Mocz_2017,Painter_2024}. It then naturally generates RC.

 The variation of density profiles with scale is presented in Figure \ref{fig:1}. The Gnedin-Primack cusp, Bahcall-Wolf cusp and Gondolo-Silk spike profile exhibit continuously rising densities towards the central region of GC black hole due to gravitational relaxation and black hole interactions respectively. In contrast, the Plummer profile shows a flattened central core up to $\sim 400$ au. The SIDM profiles (both $r^2$ and $r^3$ model) exhibit an isothermal core up to $\sim 1000$ au. The FDM profile shows an almost flattened central density.
  
  The variation of mass profiles with scale is presented in Figure \ref{fig:2}. The mass enclosed by Bahcall-Wolf cusp and Gondolo-Silk spike is comparatively higher at small radii, reaching significantly higher enclosed masses within S2's orbit due to their central density enhancements. In contrast, cored profiles including Plummer, SIDM ($r^2$ and $r^3$) and FDM exhibit more gradual mass growth in the inner regions due to their flattened central densities. At intermediate scales ($1000-3000$ au), the profiles transition from core/cusp dominated behavior to standard halo behavior. The FDM profile exhibits maximum mass accumulation at larger scales compared to other profiles. In the next section we present the precession induced by these different profiles.

\subsection{Dark matter induced precession in S-Star orbits}
In this section the precession of orbits induced by the different dark matter profiles is estimated. We estimate pericentre shift for every dark matter profile by using equation (\ref{eq16}). We then generate contour plot of logarithmic ratio of pericentre shift with dark matter to pericentre shift in GR (Schwarzschild effect) for semi-major axis in the range ($45-3500$) au and eccentricity in the range ($0.01-0.99$) (see Figures \ref{fig2} - \ref{fig7}). In all contour plots the dashed lines represent the negative ratios which indicate the regions where the Schwarzschild precession is dominant over dark matter precession. On the other hand, the solid lines indicate the positive ratios which highlight the regions where the dark matter precession dominates over Schwarzschild precession. The cases for all the profiles are highlighted below.

\subsubsection*{Plummer profile:}
Following the steps discussed in Section \ref{sec2}, equation (\ref{eq13}) - (\ref{eq16})
is worked out for the profile. The precession induced by this profile is given by,

\begin{equation}\label{eq30}
    \Delta\theta_{Pl}=-\frac{8 \pi \rho_o r_o^3 a^3(1-e^2)^3}{3 M_{bh}\ e}. I_{Pl}
\end{equation}

\noindent Where, the integral $I_{Pl}$ is given by,

\begin{equation}\label{eq31}
    I_{Pl}=\int_1^{-1}\frac{zdz}{\sqrt{1-z^2}\left[r_o^2(1+ez)^2+a^2(1-e^2)^2\right]^{3/2}}
\end{equation}

The integral (\ref{eq31}) is solved numerically. The retrograde precession caused by equation (\ref{eq30}) is compared against prograde precession induced by GR (Schwarzschild) given by $\Delta\theta_{Sch}=6\pi G M_{bh}/c^2 a(1-e^2)$. The contour for $\log_{10}|\Delta\theta_{Pl}/\Delta\theta_{Sch}|$ is presented in Figure \ref{fig2}.

\subsubsection*{Gnedin-Primack cusp profile:}
The precession induced by GP cusp is given by,

\begin{equation}\label{eq1aa_prec}
    \Delta\theta_{GPC}=-\frac{16 \pi \rho_o r_{bh}^{3/2}a(1-e^2)}{3 M_{bh}\  e}I_{GPC}
\end{equation}

where, the integral $I_{GPC}$ is given by,
\begin{equation}\label{eq1aa_integral}
    I_{GPC}=\int^{-1}_{1} \frac{zdz}{\sqrt{1-z^2}}\left[\frac{(a(1-e^2))^{1/2}}{(1+e z)^{3/2}}-\frac{l^{3/2}}{a(1-e^2)}\right]
\end{equation}

The integral (\ref{eq1aa_integral}) has an analytical form. Therefore, the final expression (\ref{eq1aa_prec}) becomes:

\begin{equation}\label{eq1aa_final}
    \Delta\theta_{GPC}=-\frac{16 \pi \rho_o r_{bh}^{3/2}a^{3/2}(1-e^2)^{3/2}}{3 M_{bh} e^2}f(e)
\end{equation}

where, $f(e)$ is purely a function of $e$ given by,

\begin{equation}\label{eq1aa_fe}
    f(e)=\frac{E\left(\frac{2e}{-1+e}\right)-(1+e)K\left(\frac{2e}{-1+e}\right)}{\sqrt{-1+e}(1+e)}+\frac{E\left(\frac{2e}{1+e}\right)-(1-e)K\left(\frac{2e}{1+e}\right)}{\sqrt{1+e}(1-e)}
\end{equation}

where, $K(x)$ and $E(x)$ are elliptic integrals of first kind and second kind respectively. 

 The retrograde precession predicted by this profile (\ref{eq1aa_final}) is compared against prograde precession induced by GR (Schwarzschild). The contour for $\log_{10}|\Delta\theta_{GPC}/\Delta\theta_{Sch}|$ is presented in Figure \ref{fig2a}.

\subsubsection*{Bahcall-Wolf cusp profile:}
The precession induced in this profile is given by,

\begin{equation}\label{eq32}
    \Delta\theta_{BWC}=-\frac{32\pi \rho_o r_o^{7/4}a^{5/4}(1-e^2)^{5/4}}{5 M_{bh}\ e}. I_{BWC}
\end{equation}

\noindent where, the integral $I_{BWC}$ is given by,

\begin{equation}\label{eq33}
    I_{BWC}=\int_{1}^{-1}\frac{zdz}{(1+ez)^{5/4}\sqrt{1-z^2}}
\end{equation}

The integral (\ref{eq33}) has an analytical form. Hence, the final expression (\ref{eq32}) becomes,

\begin{equation}\label{eq34}
    \Delta\theta_{BWC}=-\frac{32 \pi \rho_o r_o^{7/4}a^{5/4}(1-e^2)^{1/4}}{5 M_{bh} e^2 (1+\sqrt{1-e^2})^{1/4}}.g(e)
\end{equation}

\noindent where $g(e)$ is purely a function of $e$ given by,

\begin{multline}\label{eq35}
    g(e)=2^{11/4}[-(-(1-e^2)(2-e^2+2\sqrt{1-e^2}))]^{1/4}\\E\left(\frac{1}{2}-\frac{(1-e^2)^{1/4}}{\sqrt{2}\sqrt{1+\sqrt{1-e^2}}}\right)
    -2^{5/4}[1-e^2+\sqrt{1-e^2}\\+\sqrt{2}(-(-(1-e^2)(2-e^2+2\sqrt{1-e^2})))^{1/4}]\\K\left(\frac{1}{2}-\frac{(1-e^2)^{1/4}}{\sqrt{2}\sqrt{1+\sqrt{1-e^2}}}\right)
\end{multline}
    
where, $K(x)$ and $E(x)$ are elliptic integrals of first kind and second kind respectively.

The retrograde precession predicted by Bahcall-Wolf profile is compared with prograde precession induced by GR (Schwarzschild) with the help of the contour plot of $\log_{10}|\Delta\theta_{BWC}/\Delta\theta_{Sch}|$ presented in Figure \ref{fig3}.

\subsubsection*{Gondolo-Silk profile:}
The precession induced by the GS spike is given by,

\begin{equation}\label{eq36}
    \Delta\theta_{GS}=-\frac{8\pi \rho_{halo}(R_{sp})R_{sp}^{\gamma_{sp}}a(1-e^2)}{(3-\gamma_{sp})M_{bh}\ e}. I_{GS}
\end{equation}

\noindent where, the integral $I_{GS}$ is given by,

\begin{equation}\label{eq37}
    I_{GS}=\int_{1}^{-1}\frac{zdz}{\sqrt{1-z^2}}\left[\frac{[a(1-e^2)]^{2-\gamma_{sp}}}{(1+ez)^{3-\gamma_{sp}}}-\frac{(2R_{sch})^{3-\gamma_{sp}}}{a(1-e^2)}\right]
\end{equation}

The integral (\ref{eq37}) has an analytical solution. Equation (\ref{eq36}) finally takes the form,

\begin{multline}\label{eq38}
    \Delta\theta_{GS}=-\frac{4\pi^2\rho_{halo}(R_{sp})R_{sp}^{\gamma_{sp}}[a(1-e^2)]^{3-\gamma_{sp}}}{M_{bh}}\\ \,{}_2F_1\left(\frac{5-\gamma_{sp}}{2},2-\frac{\gamma_{sp}}{2};2;e^2\right)
\end{multline}

Where, ${}_2F_1(a,b;c;z)$ is the hypergeometric function. Equation (\ref{eq38}) is the exact analytical form for the precession induced by GS spike profile. The retrograde precession predicted by this profile is compared with prograde precession induced by GR (Schwarzschild). The contour for $\log_{10}|\Delta\theta_{GS}/\Delta\theta_{Sch}|$ is presented in Figure \ref{fig4}.

\subsubsection*{SIDM ($r^2$) profile:}
The precession induced by the SIDM ($r^2$) profile is given by,

\begin{equation}\label{eq39}
    \Delta\theta_{SIDM(r^2)}=-\frac{8\pi \rho_o r_o^2}{M_{bh}\ e}.I_{SIDM(r^2)}
\end{equation}

\noindent where, the integral $I_{SIDM(r^2)}$ is expressed as,

\begin{equation}\label{eq40}
    I_{SIDM(R^2)}=\int_{1}^{-1}\frac{zdz}{\sqrt{1-z^2}}\left[\frac{a(1-e^2)-r_o(1+ez)\tan^{-1}\left(\frac{a(1-e^2)}{r_o(1+ez)}\right)}{(1+ez)}\right]
\end{equation}

The integral (\ref{eq40}) has no analytical closed form and is solved numerically. The retrograde precession predicted by this profile is compared with prograde precession induced by GR (Schwarzschild) with the help of contour plot of $\log_{10}|\Delta\theta_{SIDM(r^2)}/\Delta\theta_{Sch}|$ presented in Figure \ref{fig5}.

\subsubsection*{SIDM ($r^3$) profile:}
The precession induced by SIDM ($r^3$) profile is given by,

\begin{equation}\label{eq41}
    \Delta\theta_{SIDM (r^3)}=-\frac{8\pi\rho_o r_o^3}{3M_{bh}\ e}.I_{SIDM(r^3)}
\end{equation}

\noindent where, the integral $I_{SIDM(r^3)}$ is given by,

\begin{equation}\label{eq42}
    I_{SIDM(R^3)}=\int_{1}^{-1}\frac{zdz}{\sqrt{1-z^2}}\log\left[1+\left(\frac{a(1-e^2)}{r_o(1+ez)}\right)^3\right]
\end{equation}

The integral (\ref{eq42}) has also no analytical closed form and is solved numerically. The retrograde precession calculated from equation (\ref{eq41}) is compared against prograde precession induced by GR (Schwarzschild). The contour for $\log_{10}|\Delta\theta_{SIDM(r^3)}/\Delta\theta_{Sch}|$ is presented in Figure \ref{fig6}.

\subsubsection*{FDM Profile:}
The precession induced by the FDM profile is given by,

\begin{equation}\label{eq43}
    \Delta\theta_{FDM}=-\frac{8\pi\rho_oa(1-e^2)}{215040 A^{3/2}M_{bh}\ e}.I_{FDM}
\end{equation}

\noindent where the integral $I_{FDM}$ is given by,

\begin{equation}\label{eq44}
    I_{FDM}=\int_{1}^{-1}\frac{zdz}{\sqrt{1-z^2}}f(a,e,z)
\end{equation}

\noindent where the function $f(a,e,z)$ is given by,

\begin{multline}
    f(a,e,z)=12A^{3/2}L^2(11895+\frac{3465A^5L^{10}}{(1+ez)^{10}}+\frac{19635A^4L^8}{(1+ez)^8}\\
    +\frac{45738A^3L^6}{(1+ez)^6}+\frac{55638A^2L^4}{(1+ez)^4}+\frac{36685AL^2}{(1+ez)^2})/{(1+ez)^3\left(1+\frac{AL^2}{(1+ez)^2}\right)^7}\\
    -\frac{3465\sqrt{A}(1+ez)}{AL^2+(1+ez)^8}-\frac{22A^{3/2}L^2}{(1+ez)^{11}\left(1+\frac{AL^2}{(1+ez)^2}\right)^6}(1575A^4L^8\\+7140A^3L^6(1+ez)^2+12474A^2L^4(1+ez)^4\\+10116AL^2(1+ez)^6+3335(1+ez)^8)+\frac{3465\tan^{-1}\left(\frac{\sqrt{A}L}{1+ez}\right)}{L}
\end{multline}

Here, $L=a(1-e^2)$ is the semi-latus rectum of the orbit. The integral (\ref{eq44}) is solved numerically. The retrograde precession calculated from equation (\ref{eq43}) is compared with prograde precession induced by GR (Schwarzschild). The contour for $|\Delta\theta_{FDM}/\Delta\theta_{Sch}|$ is presented in Figure \ref{fig7}.

\begin{figure*}
    \centering
    \includegraphics[width=\linewidth]{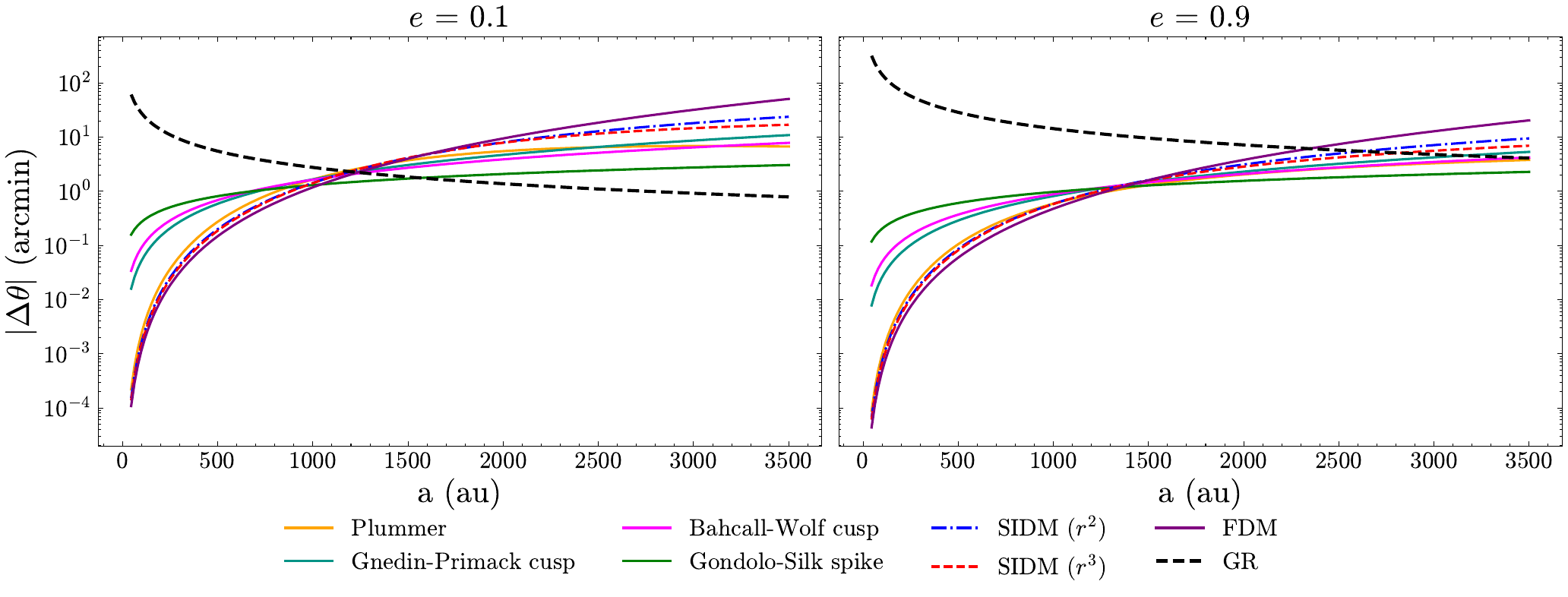}
    \caption{Variation of precession angles for different dark matter profiles as well as GR against semi-major axis.}
    \label{fig8}
\end{figure*}

For all the seven profiles considered for this study, the retrograde precession is compared with Schwarzschild precession (prograde) for minimum ($0.1$) and maximum ($0.9$) eccentricity orbits. The explicit dependence of precession angle on semi-major axis for all these cases is shown in Figure \ref{fig8} for these two eccentricities (see left and right panels of Figure \ref{fig8}). Pericentre shift of the S-star orbits calculated for all the dark matter profiles as well as the Schwarzschild pericentre shift are presented in Table \ref{tabB}.

\begin{table*}
\centering
\caption{Precession values in arc minute for different dark matter profiles considered in this work along with Schwarzschild precession.}
\label{tabB}
\resizebox{\textwidth}{!}{%
\begin{tabular}{ccccccccc}
\hline
Star &
  \begin{tabular}[c]{@{}c@{}}GR Precession\\ $(\Delta \theta _{Sch})$\end{tabular} &
  \begin{tabular}[c]{@{}c@{}}Plummer Precession\\ $(\Delta \theta_{Pl})$\end{tabular} &
  \begin{tabular}[c]{@{}c@{}}GPC Precession\\ $(\Delta \theta_{GPC})$\end{tabular} &
  \begin{tabular}[c]{@{}c@{}}BWC Precession\\ $(\Delta \theta_{BWC})$\end{tabular} &
  \begin{tabular}[c]{@{}c@{}}GS Precession\\ $(\Delta \theta_{GS})$\end{tabular} &
  \begin{tabular}[c]{@{}c@{}}SIDM ($r^2$) Precession\\ $(\Delta\theta_{SIDM(r^2)})$\end{tabular} &
  \begin{tabular}[c]{@{}c@{}}SIDM ($r^3$) Precession\\ $(\Delta \theta_{SIDM(r^3)})$\end{tabular} &
  \begin{tabular}[c]{@{}c@{}}FDM Precession\\ $(\Delta\theta_{FDM})$\end{tabular} \\ \hline
S2 &
  $+ 12.1087$ &
  \begin{tabular}[c]{@{}c@{}}$-0.6874$\\ ($5.67 \%$ of $\Delta\theta_{Sch}$)\end{tabular} &
  \begin{tabular}[c]{@{}c@{}}$-0.9117$\\ ($7.53 \%$ of $\Delta\theta_{Sch}$)\end{tabular} &
  \begin{tabular}[c]{@{}c@{}}$-0.9715$\\ ($8.02 \%$ of $\Delta\theta_{Sch}$)\end{tabular} &
  \begin{tabular}[c]{@{}c@{}}$-1.0294$\\ ($8.50 \%$ of $\Delta\theta_{Sch}$)\end{tabular} &
  \begin{tabular}[c]{@{}c@{}}$-0.6838$\\ ($5.64 \%$ of $\Delta\theta_{Sch}$)\end{tabular} &
  \begin{tabular}[c]{@{}c@{}}$-0.6781$\\ ($5.60 \%$ of $\Delta\theta_{Sch}$)\end{tabular} &
  \begin{tabular}[c]{@{}c@{}}$-0.5659$\\ ($4.67 \%$ of $\Delta\theta_{Sch}$)\end{tabular} \\
S21 &
  $+ 3.6031$ &
  \begin{tabular}[c]{@{}c@{}}$-2.8533$\\ ($79.19 \%$ of $\Delta\theta_{Sch}$)\end{tabular} &
  \begin{tabular}[c]{@{}c@{}}$-2.8325$\\ ($78.61 \%$ of $\Delta\theta_{Sch}$)\end{tabular} &
  \begin{tabular}[c]{@{}c@{}}$-2.5361$\\ ($70.39 \%$ of $\Delta\theta_{Sch}$)\end{tabular} &
  \begin{tabular}[c]{@{}c@{}}$-1.7225$\\ ($47.81 \%$ of $\Delta\theta_{Sch}$)\end{tabular} &
  \begin{tabular}[c]{@{}c@{}}$-3.8595$\\ ($107.11 \%$ of  $\Delta\theta_{Sch}$)\end{tabular} &
  \begin{tabular}[c]{@{}c@{}}$-3.6747$\\ ($101.98 \%$ of $\Delta\theta_{Sch}$)\end{tabular} &
  \begin{tabular}[c]{@{}c@{}}$-4.4222$\\ ($122.73 \%$ of $\Delta\theta_{Sch}$)\end{tabular} \\
S23 &
  $+ 1.8916$ &
  \begin{tabular}[c]{@{}c@{}}$-4.696$\\ ($248.25 \%$ of $\Delta\theta_{Sch}$)\end{tabular} &
  \begin{tabular}[c]{@{}c@{}}$-4.3538$\\ ($230.17 \%$ of $\Delta\theta_{Sch}$)\end{tabular} &
  \begin{tabular}[c]{@{}c@{}}$-3.6485$\\ ($192.87 \%$ of $\Delta\theta_{Sch}$)\end{tabular} &
  \begin{tabular}[c]{@{}c@{}}$-2.0606$\\ ($108.93 \%$ of $\Delta\theta_{Sch}$)\end{tabular} &
  \begin{tabular}[c]{@{}c@{}}$-7.1124$\\ ($375.99 \%$ of $\Delta\theta_{Sch}$)\end{tabular} &
  \begin{tabular}[c]{@{}c@{}}$-6.6827$\\ ($353.28 \%$ of $\Delta\theta_{Sch}$)\end{tabular} &
  \begin{tabular}[c]{@{}c@{}}$-8.9598$\\ ($473.66 \%$ of $\Delta\theta_{Sch}$)\end{tabular} \\
S29 &
  $+13.7035$ &
  \begin{tabular}[c]{@{}c@{}}$-1.8905$\\ ($13.79 \%$ of $\Delta\theta_{Sch}$)\end{tabular} &
  \begin{tabular}[c]{@{}c@{}}$-2.8044$\\ ($20.47 \%$ of $\Delta\theta_{Sch}$)\end{tabular} &
  \begin{tabular}[c]{@{}c@{}}$-2.3533$\\ ($17.17 \%$ of $\Delta\theta_{Sch}$)\end{tabular} &
  \begin{tabular}[c]{@{}c@{}}$-1.6034$\\ ($11.7 \%$ of $\Delta\theta_{Sch}$)\end{tabular} &
  \begin{tabular}[c]{@{}c@{}}$-4.5648$\\ ($33.31 \%$ of $\Delta\theta_{Sch}$)\end{tabular} &
  \begin{tabular}[c]{@{}c@{}}$-3.4420$\\ ($25.11 \%$ of $\Delta\theta_{Sch}$)\end{tabular} &
  \begin{tabular}[c]{@{}c@{}}$-8.0593$\\ ($58.81 \%$ of $\Delta\theta_{Sch}$)\end{tabular} \\
S38 &
  $+6.97002$ &
  \begin{tabular}[c]{@{}c@{}}$-1.1692$\\ ($16.77 \%$ of $\Delta\theta_{Sch}$)\end{tabular} &
  \begin{tabular}[c]{@{}c@{}}$-1.3442$\\ ($19.29 \%$ of $\Delta\theta_{Sch}$)\end{tabular} &
  \begin{tabular}[c]{@{}c@{}}$-1.3568$\\ ($19.46 \%$ of $\Delta\theta_{Sch}$)\end{tabular} &
  \begin{tabular}[c]{@{}c@{}}$-1.2315$\\ ($17.66 \%$ of $\Delta\theta_{Sch}$)\end{tabular} &
  \begin{tabular}[c]{@{}c@{}}$-1.2053$\\ ($17.29 \%$ of $\Delta\theta_{Sch}$)\end{tabular} &
  \begin{tabular}[c]{@{}c@{}}$-1.1971$\\ ($17.17 \%$ of $\Delta\theta_{Sch}$)\end{tabular} &
  \begin{tabular}[c]{@{}c@{}}$-1.0722$\\ ($15.38 \%$ of $\Delta\theta_{Sch}$)\end{tabular} \\
S55 &
  $+6.7424$ &
  \begin{tabular}[c]{@{}c@{}}$-0.7259$\\ ($10.76 \%$ of $\Delta\theta_{Sch}$)\end{tabular} &
  \begin{tabular}[c]{@{}c@{}}$-0.9779$\\ ($14.50 \%$ of $\Delta\theta_{Sch}$)\end{tabular} &
  \begin{tabular}[c]{@{}c@{}}$-1.0472$\\ ($15.53 \%$ of $\Delta\theta_{Sch}$)\end{tabular} &
  \begin{tabular}[c]{@{}c@{}}$-1.0611$\\ ($15.73 \%$ of $\Delta\theta_{Sch}$)\end{tabular} &
  \begin{tabular}[c]{@{}c@{}}$-0.6339$\\ ($9.40 \%$ of $\Delta\theta_{Sch}$)\end{tabular} &
  \begin{tabular}[c]{@{}c@{}}$-0.6227$\\ ($9.24 \%$ of $\Delta\theta_{Sch}$)\end{tabular} &
  \begin{tabular}[c]{@{}c@{}}$-0.5082$\\ ($7.53 \%$ of $\Delta\theta_{Sch}$)\end{tabular} \\
S60 &
  $+1.7484$ &
  \begin{tabular}[c]{@{}c@{}}$-5.6980$\\ ($325.89 \%$ of $\Delta\theta_{Sch}$)\end{tabular} &
  \begin{tabular}[c]{@{}c@{}}$-7.1248$\\ ($407.50 \%$ of $\Delta\theta_{Sch}$)\end{tabular} &
  \begin{tabular}[c]{@{}c@{}}$-5.4823$\\ ($313.56 \%$ of $\Delta\theta_{Sch}$)\end{tabular} &
  \begin{tabular}[c]{@{}c@{}}$-2.6083$\\ ($149.18 \%$ of $\Delta\theta_{Sch}$)\end{tabular} &
  \begin{tabular}[c]{@{}c@{}}$-13.5998$\\ ($777.84 \%$ of $\Delta\theta_{Sch}$)\end{tabular} &
  \begin{tabular}[c]{@{}c@{}}$-10.5278$\\ ($602.14 \%$ of $\Delta\theta_{Sch}$)\end{tabular} &
  \begin{tabular}[c]{@{}c@{}}$-26.6739$\\ ($1525.62 \%$ of $\Delta\theta_{Sch}$)\end{tabular} \\
S62 &
  $+77.4588$ &
  \begin{tabular}[c]{@{}c@{}}$-0.1248$\\ ($0.16 \%$ of $\Delta\theta_{Sch}$)\end{tabular} &
  \begin{tabular}[c]{@{}c@{}}$-0.2718$\\ ($0.35 \%$ of $\Delta\theta_{Sch}$)\end{tabular} &
  \begin{tabular}[c]{@{}c@{}}$-0.3319$\\ ($0.42 \%$ of $\Delta\theta_{Sch}$)\end{tabular} &
  \begin{tabular}[c]{@{}c@{}}$-0.5438$\\ ($0.70 \%$ of $\Delta\theta_{Sch}$)\end{tabular} &
  \begin{tabular}[c]{@{}c@{}}$-0.1293$\\ ($0.16 \%$ of $\Delta\theta_{Sch}$)\end{tabular} &
  \begin{tabular}[c]{@{}c@{}}$-0.1263$\\ ($0.16 \%$ of $\Delta\theta_{Sch}$)\end{tabular} &
  \begin{tabular}[c]{@{}c@{}}$-0.0812$\\ ($0.10 \%$ of $\Delta\theta_{Sch}$)\end{tabular} \\
S4711 &
  $+10.7039$ &
  \begin{tabular}[c]{@{}c@{}}$-0.2924$\\ ($2.73 \%$ of $\Delta\theta_{Sch}$)\end{tabular} &
  \begin{tabular}[c]{@{}c@{}}$-0.5620$\\ ($0.09 \%$ of $\Delta\theta_{Sch}$)\end{tabular} &
  \begin{tabular}[c]{@{}c@{}}$-0.6587$\\ ($6.15 \%$ of $\Delta\theta_{Sch}$)\end{tabular} &
  \begin{tabular}[c]{@{}c@{}}$-0.8276$\\ ($7.73 \%$ of $\Delta\theta_{Sch}$)\end{tabular} &
  \begin{tabular}[c]{@{}c@{}}$-0.2347$\\ ($2.19 \%$ of $\Delta\theta_{Sch}$)\end{tabular} &
  \begin{tabular}[c]{@{}c@{}}$-0.2252$\\ ($2.10 \%$ of $\Delta\theta_{Sch}$)\end{tabular} &
  \begin{tabular}[c]{@{}c@{}}$-0.1749$\\ ($1.63 \%$ of $\Delta\theta_{Sch}$)\end{tabular} \\
S4713 &
  $+1.8753$ &
  \begin{tabular}[c]{@{}c@{}}$-4.0204$\\ ($214.38 \%$ of $\Delta\theta_{Sch}$)\end{tabular} &
  \begin{tabular}[c]{@{}c@{}}$-3.3817$\\ ($180.33 \%$ of $\Delta\theta_{Sch}$)\end{tabular} &
  \begin{tabular}[c]{@{}c@{}}$-2.9578$\\ ($157.72 \%$ of $\Delta\theta_{Sch}$)\end{tabular} &
  \begin{tabular}[c]{@{}c@{}}$-1.8142$\\ ($96.74 \%$ of $\Delta\theta_{Sch}$)\end{tabular} &
  \begin{tabular}[c]{@{}c@{}}$-4.8328$\\ ($257.71 \%$ of $\Delta\theta_{Sch}$)\end{tabular} &
  \begin{tabular}[c]{@{}c@{}}$-4.8363$\\ ($257.89 \%$ of $\Delta\theta_{Sch}$)\end{tabular} &
  \begin{tabular}[c]{@{}c@{}}$-5.0351$\\ ($268.49 \%$ of $\Delta\theta_{Sch}$)\end{tabular} \\
S4714 &
  $+108.522$ &
  \begin{tabular}[c]{@{}c@{}}$-0.1235$\\ ($0.11 \%$ of $\Delta\theta_{Sch}$)\end{tabular} &
  \begin{tabular}[c]{@{}c@{}}$-0.2631$\\ ($0.24 \%$ of $\Delta\theta_{Sch}$)\end{tabular} &
  \begin{tabular}[c]{@{}c@{}}$-0.3145$\\ ($0.29 \%$ of $\Delta\theta_{Sch}$)\end{tabular} &
  \begin{tabular}[c]{@{}c@{}}$-0.5104$\\ ($0.47 \%$ of $\Delta\theta_{Sch}$)\end{tabular} &
  \begin{tabular}[c]{@{}c@{}}$-0.1453$\\ ($0.13 \%$ of $\Delta\theta_{Sch}$)\end{tabular} &
  \begin{tabular}[c]{@{}c@{}}$-0.1431$\\ ($0.13 \%$ of $\Delta\theta_{Sch}$)\end{tabular} &
  \begin{tabular}[c]{@{}c@{}}$-0.083$\\ ($0.07 \%$ of $\Delta\theta_{Sch}$)\end{tabular} \\
S4715 &
  $+2.4386$ &
  \begin{tabular}[c]{@{}c@{}}$-2.3314$\\ ($95.60 \%$ of $\Delta\theta_{Sch}$)\end{tabular} &
  \begin{tabular}[c]{@{}c@{}}$-2.1163$\\ ($86.78 \%$ of $\Delta\theta_{Sch}$)\end{tabular} &
  \begin{tabular}[c]{@{}c@{}}$-2.0013$\\ ($82.06 \%$ of $\Delta\theta_{Sch}$)\end{tabular} &
  \begin{tabular}[c]{@{}c@{}}$-1.4607$\\ ($59.89 \%$ of $\Delta\theta_{Sch}$)\end{tabular} &
  \begin{tabular}[c]{@{}c@{}}$-2.2003$\\ ($90.23 \%$ of $\Delta\theta_{Sch}$)\end{tabular} &
  \begin{tabular}[c]{@{}c@{}}$-2.2101$\\ ($90.62 \%$ of $\Delta\theta_{Sch}$)\end{tabular} &
  \begin{tabular}[c]{@{}c@{}}$-1.9341$\\ ($79.31 \%$ of $\Delta\theta_{Sch}$)\end{tabular} \\
S4716 &
  $+15.8564$ &
  \begin{tabular}[c]{@{}c@{}}$-0.0906$\\ ($0.57 \%$ of $\Delta\theta_{Sch}$)\end{tabular} &
  \begin{tabular}[c]{@{}c@{}}$-0.2975$\\ ($1.88 \%$ of $\Delta\theta_{Sch}$)\end{tabular} &
  \begin{tabular}[c]{@{}c@{}}$-0.3880$\\ ($2.45 \%$ of $\Delta\theta_{Sch}$)\end{tabular} &
  \begin{tabular}[c]{@{}c@{}}$-0.6198$\\ ($3.91 \%$ of $\Delta\theta_{Sch}$)\end{tabular} &
  \begin{tabular}[c]{@{}c@{}}$-0.0679$\\ ($0.43 \%$ of $\Delta\theta_{Sch}$)\end{tabular} &
  \begin{tabular}[c]{@{}c@{}}$-0.0634$\\ ($0.39 \%$ of $\Delta\theta_{Sch}$)\end{tabular} &
  \begin{tabular}[c]{@{}c@{}}$-0.0483$\\ ($0.30 \%$ of $\Delta\theta_{Sch}$)\end{tabular} \\ \hline
\end{tabular}%
}
\end{table*}

\section{Prospects for detection in the orbit of S-stars}\label{sec4}
In order to investigate the detectability of dark matter induced precession we consider orbits of S2, S29, S38 and S55 which crossed their periapsis by 2022 \citep{GRAVITY2022,GRAVITY2024}. We also consider the orbits of some other stars including S4711, S4713-S4716, S21, S23, S29, S60, S62 which are expected to reach their periapsis within the next decade \citep{Gillessen_2017,Peißker_2020a,Peißker_2020b,Peißker_2022}. The orbital data for all the stars are presented in Table \ref{tabA}. In the following subsections, we investigate the possibility of detection of these effects in the orbit of S-stars. In subsection \ref{sec41}, we consider the orbit of S2, take help of the $f_{sp}$ parameter (deviation from Schwarzschild precession) and examine detectability of dark matter effect. 

Here, we note that a modification to the theory of gravity near the black hole can affect the general relativistic and dark matter induced precession. Therefore, we take into account two well studied alternatives of GR to see the effect on the $f_{sp}$ values. These are $f(R)$ gravity theory and Scalar-Tensor-Vector Gravity theory which have exact analytical expressions for Schwarzschild like precession \citep{Paul_2024,STVG_Della_Monica_2023b}. In subsection \ref{sec42}, we estimate the astrometric shift of the pericentres of other S-stars and compare with astrometric capabilities of existing and upcoming telescopes.

\subsection{The orbit of S2}\label{sec41}
The orbit of S2 remains the most precisely tracked orbit till date. \cite{Gravity_2020} measured the Schwarzschild precession of S2's orbit with $5\sigma$ accuracy. This measurement has been further improved by \cite{GRAVITY2022} and \cite{GRAVITY2024}. A parameter $f_{sp}$ has been introduced which is related to total precession ($\Delta\theta$) as

\begin{equation}\label{eq45}
    \Delta\theta=f_{sp}\Delta\theta_{Sch}
\end{equation}

\noindent Here, $\Delta\theta_{Sch}$ is the Schwarzschild precession. When $f_{sp}=0$, it is purely Newtonian gravity. $f_{sp}=1$ corresponds to purely Schwarzschild precession. This can be expressed as \citep{Paul_2024}

\begin{equation}\label{eq46}
    f_{sp}=1+\left[\frac{\Delta\theta_i-\Delta\theta_{Sch}}{\Delta\theta_{Sch}}\right]
\end{equation}

where, $\Delta\theta_i=\Delta\theta_{Sch}+\Delta\theta_{X}$ is the total precession and $\Delta\theta_{X}$ is the contribution from different mass profiles. The parameter $f_{sp}$ was initially constrained as $f_{sp}=1.10 \pm 0.19\ (5\sigma)$ \citep{Gravity_2020} and has recently been refined as $f_{sp}=1.135\pm 0.110\ (10\sigma)$ \citep{GRAVITY2024}. We estimate the values of $f_{sp}$ corresponding to the orbital precession of S2 in the presence of various profiles (i.e., Schwarzschild plus dark matter profiles) and summarize them in Table \ref{tab2}. We find that all of the estimated values of $f_{sp}$ fall outside the most recent observational constraint. Nevertheless, the updated allowed interval $f_{sp}\in (1.025-1.245)$ suggests the presence of additional prograde precession rather than retrograde contributions. This motivates us to explore possible corrections arising from modified theories of gravity. Let us see the effect of $f(R)$ gravity theory and STVG theory.

\begin{table}
\centering
\caption{Estimated $f_{sp}$ values for different profiles}
\label{tab2}
\resizebox{0.3\textwidth}{!}{%
\begin{tabular}{cc}
\hline
Model                                           & $f_{sp}$ Value \\ \hline
$\Delta\theta_{Sch} + \Delta\theta_{Pl}$        & $0.9430$       \\
$\Delta\theta_{Sch} + \Delta\theta_{GPC}$        & $0.9247$       \\
$\Delta\theta_{Sch} + \Delta\theta_{BWC}$       & $0.9201$       \\
$\Delta\theta_{Sch} + \Delta\theta_{GS}$        & $0.9149$       \\
$\Delta\theta_{Sch} + \Delta\theta_{SIDM(r^2)}$ & $0.9435$       \\
$\Delta\theta_{Sch} + \Delta\theta_{SIDM(r^3)}$ & $0.9436$       \\
$\Delta\theta_{Sch} + \Delta\theta_{FDM}$       & $0.9512$       \\ \hline
\end{tabular}%
}
\end{table}

\subsubsection*{$f(R)$ gravity:}
The $f(R)$ theory of gravity is a modified gravity model that is used to address accelerated expansion as well as flat rotation curves without incorporating dark energy and dark matter respectively \citep{Capozziello_2002,Capozziello_2003,Carroll_2004}. In this theory $f(R)$ is a function of the curvature scalar $R$. $f(R)=R$ corresponds to GR. In other theories the derivative $df(R)/dR$ acts as a scalar field, $\psi$. This is the additional gravitational mode with the space time metric field. This is known as scalaron. Effect of the scalaron field on the precession of the compact stellar orbits near the GC black hole is studied earlier by \cite{Kalita2020,Kalita_2021,Paul_2023} and \cite{Paul_2024}. \cite{Paul_2024} calculated the exact Schwarzschild like precession in $f(R)$ gravity theory (for a spherically symmetric and static solution of $f(R)$ gravity field equation) and is expressed as

\begin{multline}\label{eq47}
    (\Delta\theta)_{SchS}=\frac{6\pi G M_{bh}}{c^2 \psi_o a(1-e^2)}\left(1+\frac{1}{3}e^{-M_\psi a(1-e^2)}\right)\\+\frac{4\pi G M_{bh} M_\psi}{3 c^2 \psi_o}e^{-M_\psi a(1-e^2)}
    +\frac{2\pi a^2(1-e^2) M_\psi^2}{6}e^{-M_\psi a(1-e^2)}\\+\frac{2\pi G M_{bh} a(1-e^2) M_\psi^2}{6 c^2 \psi_o}e^{-M_\psi a(1-e^2)}
\end{multline}

Here, $\psi_o$ is the scalaron field amplitude ($\psi_o=(df(R)/dR)_{R_o}$, $R_o$ being the background scalar curvature) and $M_\psi$ is the mass of the scalaron. The $f(R)$ theory naturally introduces a Yukawa correction to the Newtonian gravitational potential \citep{Kalita2018}. The parameter $\psi_o$ is related to the Yukawa coupling constant $\alpha$ as $\psi_o=1/3\alpha$ \citep{Kalita2020}. \cite{GRAVITY2025} constrained $\alpha$ around $0.0031$ which corresponds to $\psi_o=107.527$. The scalaron mass is generally a free parameter whose influence on stellar orbits and black hole shadow have been studied earlier by \cite{Kalita2020,Paul_2023,Kalita_2023} and \cite{Paul_2024}. These authors converged to the conclusion that scalarons which are heavier than the graviton mass ($\sim 10^{-23}$ eV) bounded by LIGO-Virgo gravitational wave observation \citep{LIGO_Virgo_2016} are generally compatible with stellar orbits and black hole shadow measurements. Here, we derive that the scalaron mass, $M_\psi=4.45 \times 10^{-19}$ eV produces Schwarzschild like precession which corresponds to $f_{sp}=1.111$. It comfortably falls in the updated range of $f_{sp}$ as mentioned above.

\subsubsection*{Scalar-Tensor-Vector Gravity:}
This class of gravity theory has been used to fit rotation curves and large scale structure (LSS) data \citep{Moffat_2006}. The first order precession in this theory is given by \citep{STVG_Della_Monica_2023b}

\begin{equation}\label{eq48}
    \Delta\theta_{STVG}=\left(1+\frac{5}{6}\alpha' \right)\frac{6\pi G_N M_{bh}}{c^2 a (1-e^2)}
\end{equation}

where, $\alpha'$ is a parameter of the theory and the Newtonian gravitational constant $G_N$ is connected to its counterpart $G$ of the theory as $G=G_N(1+\alpha')$. For $\alpha'=0$ the theory reduces to GR. \cite{STVG_Della_Monica} constrained the parameter $\alpha'$ as $\alpha'\lesssim 0.662$ with $99.7\%$ confidence limit. The relation between $\alpha'$ and $f_{sp}$ is found as
\begin{equation}
    \alpha'=\frac{6}{5}(f_{sp}-1)
\end{equation}

Therefore, the updated bound on $f_{sp}$ corresponds to $\alpha' \in (0.03-0.294)$. We note with interest that \cite{Moffat_2008} predicted $\alpha'=0.03$ from the analysis of globular clusters with masses above $10^6\ M_{\odot}$.

\begin{figure}
    \centering
    \includegraphics[width=\linewidth]{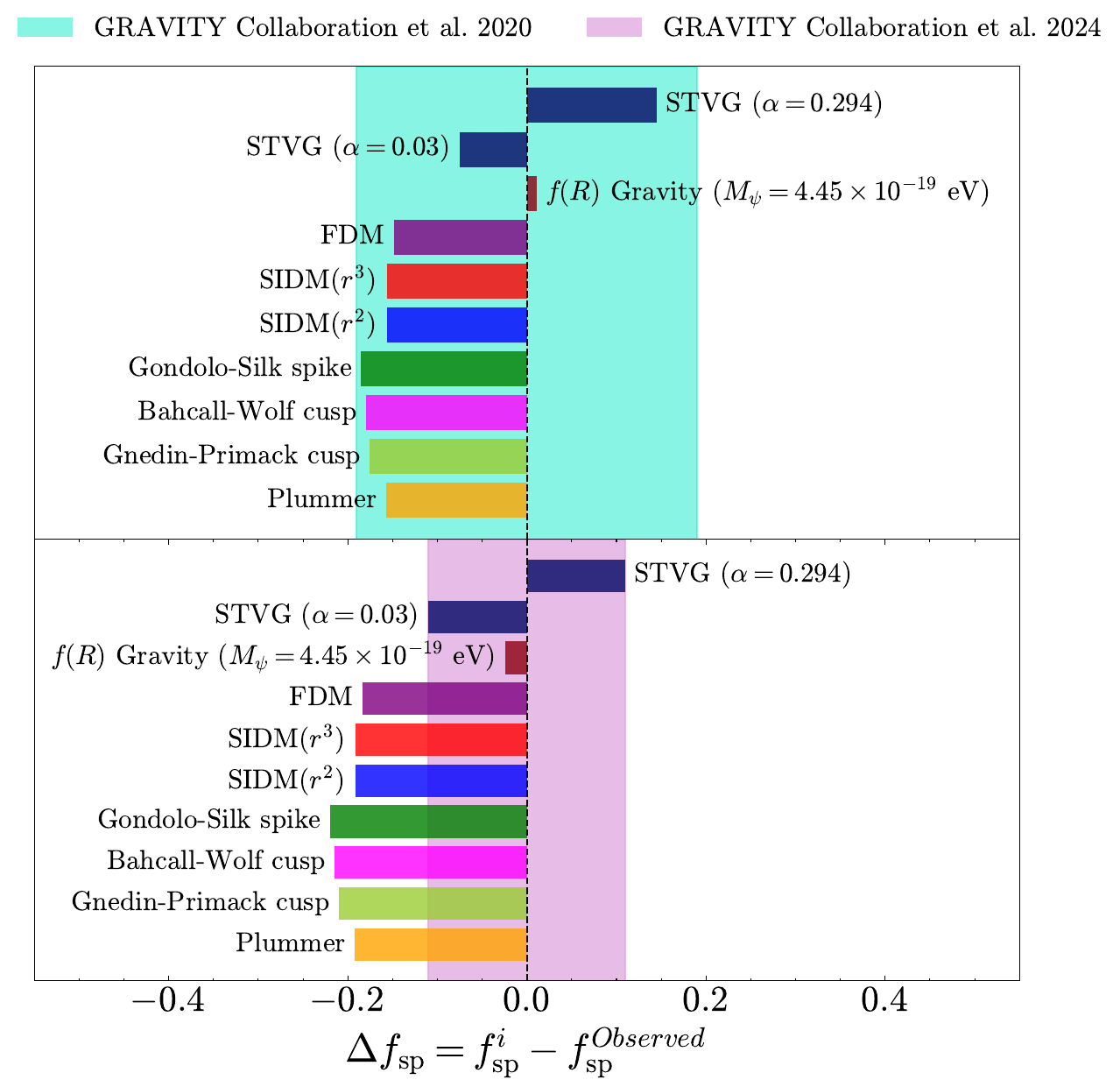}
    \caption{The deviation parameter $f_{sp}$ for different dark matter models as well as modified gravity models}
    \label{fig10}
\end{figure}

We compare the estimated $f_{sp}$ values for different dark matter profiles and modified gravity theories with the old and updated observational bounds obtained by \cite{Gravity_2020,GRAVITY2024}. We estimate the theoretical error as
\begin{equation}
    \Delta f_{sp}=f_{sp}^{i}- f_{sp}^{Observed}
\end{equation}

where, $f_{sp}^{Observed}$ is the central value of the observational bounds and $f_{sp}^i$ is the value estimated by considering either GR plus dark matter effect or GR plus modified gravity effect. We compare this $\Delta f_{sp}$ against \cite{Gravity_2020} and \cite{GRAVITY2024} bound. The deviations are presented in Figure \ref{fig10}. This helps us to identify whether dark matter or modified gravity effect is prominent within the orbit of S2.

\subsection{Astrometric shift of S-stars:}\label{sec42}
The existing Keck telescope, the GRAVITY instrument at the VLT and the upcoming Thirty Meter Telescope (TMT) provide the necessary precision to probe gravitational effects and matter distributions in the vicinity of the GC black hole. Their astrometric accuracies are $\sigma_{Keck}=0.16$ mas, $\sigma_{GRAVITY}=0.030$ mas and $\sigma_{TMT}=0.015$ mas respectively
\citep{Hees_2017,Gravity_2018}. Following the approach adopted by \cite{Zakharov_2018}, we take a characteristic angle $\phi=2\sigma$ which corresponds to a $95\%$ confidence level for detectability with these instruments. Hence, the angles (in $\mu$as) up to which these facilities will be able to detect orbital shifts are: $\phi_{Keck}=320\ \mu as,\ 
    \phi_{GRAVITY}= 60\ \mu as,\ \phi_{TMT}=30 \ \mu as$. The astrometric shift of the pericentre in the sky is estimated as
    
\begin{equation}\label{eq49}
    \phi=\Delta\theta \times\frac{a}{D} \sin(i)
\end{equation}

\begin{figure*}
    \centering
    \includegraphics[width=0.9\linewidth]{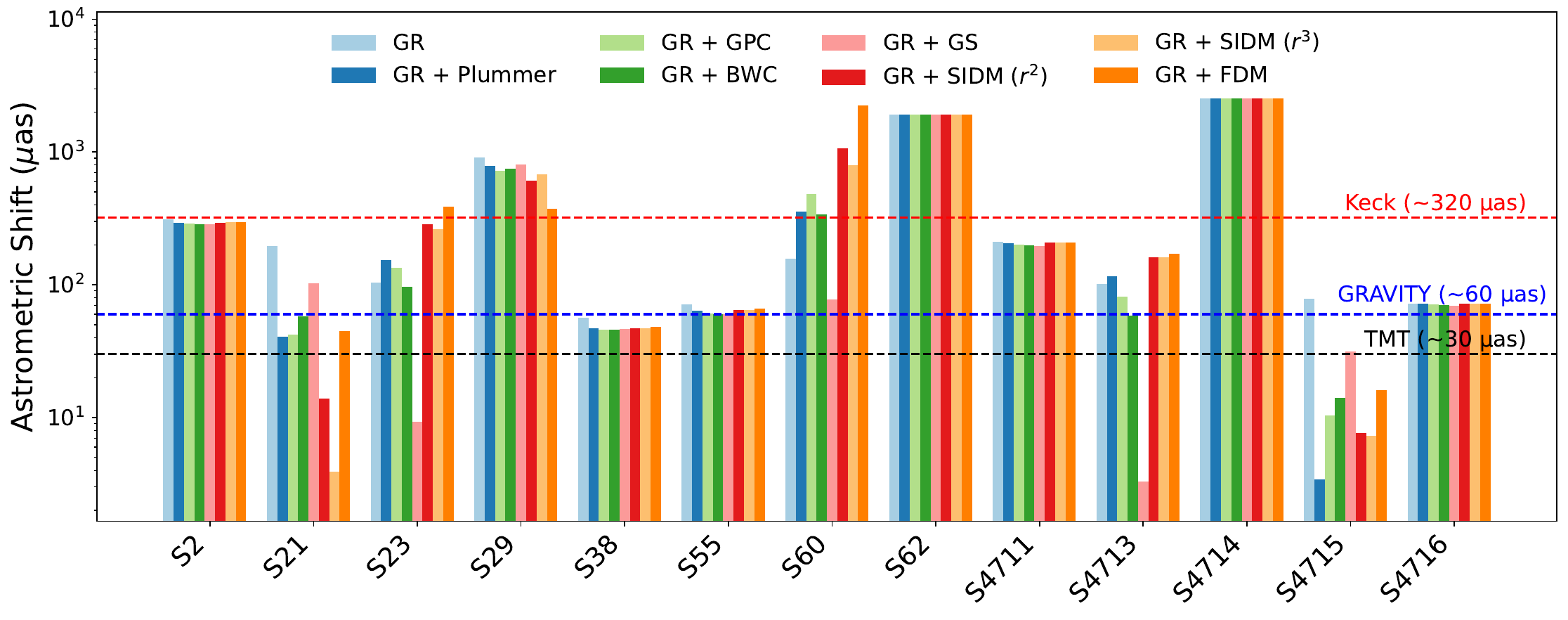}
    \caption{Astrometric shift caused by different dark matter profiles (vertical bars) along with astrometric capabilities of Keck, GRAVITY and TMT.}
    \label{fig9}
\end{figure*}

Here, $a$ is the semi-major axis, $D$ is the distance to the GC black hole and $i$ is the inclination of the orbit in degrees. Using equation (\ref{eq49}), we estimate the astrometric shift for the 13 S-stars considered in this study. Distance to the GC black hole has been taken as $D=8275.9$ pc \citep{GRAVITY2024}. We estimate the astrometric shift in pure Schwarzschild case as well as Schwarzschild plus dark matter case. The values of $\phi$ for all profiles in each star are presented by a bar plot in Figure \ref{fig9}. Table \ref{tabC} shows the numerical values of $\phi$. In Figure \ref{fig9} the dashed horizontal line represent astrometric accuracies of Keck, GRAVITY and TMT. The $\phi$ values below these lines for any dark matter profile will be inaccessible to these detectors.

\section{Results and discussion}\label{sec5}
In this study we calculate pericentre shift of some S-star orbits in presence of dark matter near the GC black hole. With the help of the $f_{sp}$ parameter we examine the competition between dark matter induced precession and modified gravity effect within the orbit of S2. We identify stellar orbits which are prone to dark matter induced precession to be detectable by existing and upcoming astrometric capabilities. The major findings are summarised below.

Table \ref{tabB} displays pericentre shift of the S-star orbits calculated for all the dark matter profiles and the Schwarzschild pericentre shift. It is seen that in the orbits of S23, S60 and S4713 the Schwarzschild precession is completely dominated by the dark matter precession due to Plummer, Gnedin-Primack cusp, Bahcall-Wolf cusp, Gondolo-Silk spike, SIDM ($r^2$ and $r^3$) and FDM profiles. The effect of these profiles are found to be $50 \%$ or more than that of the Schwarzschild precession in the orbits of S21 and S4715. The FDM precession is $58.81\%$ of Schwarzschild precession in the orbit of S29.

 Very near to the black hole, the Gondolo-Silk spike precession dominates the precession induced by other profiles. This is followed by the Bahcall-Wolf cusp precession. But all the dark matter induced precession values are small compared to the Schwarzschild precession. For low eccentricity orbits ($e=0.1$), the precession induced by these different profiles start dominating Schwarzschild precession after approximately $1200-1500$ au. For highly eccentric orbits ($e=0.9$), it is seen that the precession induced by these different profiles start dominating Schwarzschild precession towards wider orbits ($\sim 2400-3500$ au). Also, the FDM induced precession starts dominating towards the wider orbits. We note that there is no prograde precession for the pure Newtonian case. General relativistic correction can lead to either minor retrograde or prograde precession depending on the dark matter density profiles and distance from the black hole. For investigating possibility of detection of the precession angles predicted by several dark matter profiles we consider interference from new gravitational physics near the black hole. Prograde pericentre shift induced by $f(R)$ gravity scalaron and STVG theory are calculated.

 It is seen that all the dark matter profiles as well as modified gravity models fall within the \cite{Gravity_2020} bound on the $f_{sp}$ parameter. However, only the modified gravity models seem to consistent with the updated \cite{GRAVITY2024} bound (see Figure \ref{fig10}). None of the dark matter profiles are consistent with the current bound. This indicates that with the orbit of S2, it might not be possible to detect deviations due to dark matter distributions. However, S2 can serve as a reliable probe for understanding modified theories of gravity. The massive scalaron mode with $M_\psi\approx 10^{-19}$ eV seems compatible with the recent $f_{sp}$ bound. In case of STVG theory we have been able to reduce the upper bound on the deviation parameter on $\alpha' \in (0.03 -0.294)$ in relation to the earlier bound reported by \cite{STVG_Della_Monica}. This enhances prospect for testing alternatives to GR within the orbit of S2. We now highlight the dark matter profiles whose precession can be measured by present and upcoming telescope facilities. We refer to Figure \ref{fig9} to discuss this.

 It has been observed that in the orbit of S2, the deviation in astrometric shift in presence of different dark matter profiles from Schwarzschild case is not very significant. In the orbit of S21, there is noticeable deviation. The Plummer, BW cusp, GS spike and FDM can be distinguished using the GRAVITY and TMT facilities. However, SIDM (both $r^2$ and $r^3$) models are not accessible to the existing and upcoming facilities. In the orbit of S23, it has been observed that except for GS spike rest of the profiles are distinguishable using the available facilities. In the orbit of S29, it is seen that all the profiles should be distinguishable by all three facilities. In the orbit of S38, it has been observed that although TMT will not be able to distinguish between different profiles. Similarly, for S55 the facilities such as TMT and GRAVITY will not be able to distinguish between the different profiles. The orbit of S60 shows major deviation in presence of different profiles and the existing and upcoming facilities should be able to distinguish between the models. The orbit of S62 shows very high astrometric shift but the difference between the models is not significant and hence the facilities will not be able to differentiate them. The orbit of S4711 shows very little difference between the models and hence will be challenging for the facilities to distinguish between them. The orbit of S4713, shows some deviation between the models. However any effect due to GS spike will not be detectable by the facilities. The orbit of S4714 shows very high astrometric shift but the deviation between the models is negligible. Hence, none of the three facilities will be able to distinguish between the models. The orbit of S4715 shows significant deviation among the models but the amount of astrometric shift is below capabilities of all the three facilities and hence the effects might not be detectable through these facilities. Lastly, the orbit of S4716 shows negligible deviation among the models and hence will not be distinguishable through any facility. Overall, we see that the effect of dark matter is dominant in low eccentricity or wider orbits where relativistic effects are less. The effect of dark matter in highly relativistic orbits is almost negligible. We extract dark matter induced precession for all the profiles allowed by the most recent upper bound on the amount of dark mass near the GC black hole.

\begin{figure}
    \centering
    \includegraphics[width=0.9\linewidth]{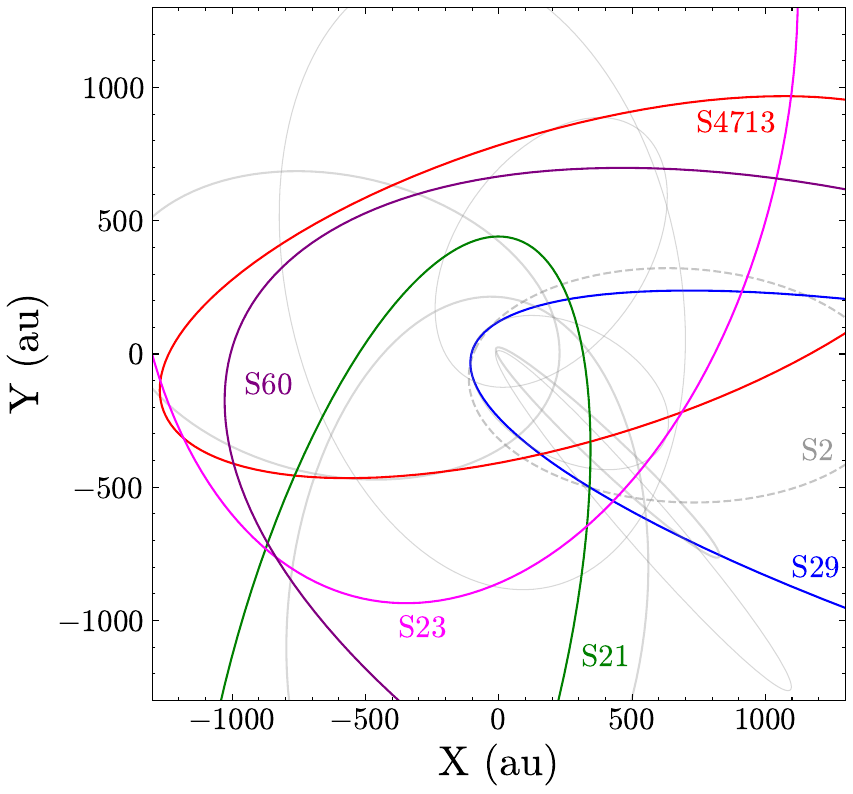}
    \caption{The projection of the S-star orbits in the sky plane investigated in this work. Potential orbits where dark matter effects could be detectable are highlighted in color. Rest of the orbits are marked in grey. The orbit of S2 is represented by dashed line.}
    \label{fig11}
\end{figure}

In Figure \ref{fig11} we display the stellar orbits where precession due to dark matter distribution can be measured by existing and upcoming astrometric facilities (see prominent orbital tracks). These orbits are generated by using the method illustrated in Sect. 3.2.5 of \cite{poisson2014gravity}. These are also orbits where some dark matter profiles can be distinguished. It is to be noted that the secular perturbation method adopted in this work is valid when the DM mass is very low compared to the black hole mass, which is perhaps realistic. However, we need a fully relativistic or Post-Newtonian (PN) treatment while dealing with extremely compact orbits embedded in extended mass distribution such as extreme mass ratio inspiral (EMRI) systems.

We conclude with the following remarks. We find that existing and upcoming facilities have the potential to distinguish between different dark matter profiles for some stars and hence they posses the capability to distinguish between possible physical processes responsible for formation of the central region of our galaxy. It may be difficult to probe or distinguish between dark matter profile by using S2's orbit. However, the orbit of S2 is a reliable probe for new gravitational physics. We produce new constraint on mass of the scalar mode in $f(R)$ gravity as $10^{-19}$ eV. The scalaron mass is strikingly similar to the Fuzzy Dark Matter particle mass constrained by \cite{ULDM2023_PRD,ULDM2023} and calculated in this work by using the upper bound on dark mass ($\sim 10^{-19}$ eV). This motivates further investigation of alternative gravity theories which promise replacement of dark matter. Low eccentricity and wider orbits are prominent probes for measuring dark matter induced precession which is accessible to present and upcoming facilities.

\section*{Acknowledgments}
DP is grateful to the Council of Scientific and Industrial Research (CSIR), Government of India, for financial support in the form of Senior Research Fellowship (SRF-Direct, File no. 09/0059(23533)/2025-EMR-I). The authors would like to thank the anonymous reviewer for helpful comments on the manuscript.

\section*{Data Availability}
All data underlying the results are presented in the figures and tables within this article.



\bibliographystyle{mnras}
\bibliography{reference} 




\appendix
\onecolumn
\section{Orbital data for S-stars}
We present the orbital data for S-stars considered in this study in Table \ref{tabA}.

\noindent
\begin{center}
\captionof{table}{Orbital parameters of the S-stars used in this work.}
\label{tabA}
\resizebox{0.8\textwidth}{!}{%
\begin{tabular}{cccccc}
\hline
Star &
$a$ (au) &
$e$ &
$i\ (^{\circ})$ &
$\omega\ (^{\circ})$ &
$\Omega\ (^{\circ})$ \\
\hline
S2$^{\mathrm{a}}$    & $1031.002 \pm 1.1887$ & $0.88444 \pm (6\times 10^{-5})$   & $134.67 \pm 0.02$ & $66.279 \pm 0.029$ & $228.21 \pm 0.03$   \\
S21$^{\mathrm{d}}$   & $1812.42 \pm 14.1945$ & $0.764 \pm 0.014$                 & $58.8\pm 1$       & $166.4 \pm 1.1$    & $259.64 \pm 0.62$   \\
S23$^{\mathrm{d}}$   & $2093.8 \pm 99.3345$  & $0.56 \pm 0.14$                   & $48 \pm 7.1$      & $39\pm 6.7$        & $249 \pm 13$        \\
S29$^{\mathrm{a}}$   & $3229.67 \pm 8.4724$  & $0.9688 \pm (9\times 10^{-5})$    & $144.24 \pm 0.09$ & $203.68 \pm 0.17$  & $4.9259 \pm 0.1590$ \\
S38$^{\mathrm{a}}$   & $1179.23 \pm 1.2693$  & $0.8181 \pm (2.2 \times 10^{-4})$ & $168.69\pm 0.19$  & $40.065\pm 1.118$  & $122.43\pm 1.12$    \\
S55$^{\mathrm{a}}$   & $862.68 \pm 0.9873$   & $0.7298 \pm (1.8 \times 10^{-4})$ & $159.59 \pm 0.17$ & $327.77 \pm 0.93$  & $319.43 \pm 0.97$   \\
S60$^{\mathrm{d}}$   & $3208.57 \pm 58.0272$ & $0.7179 \pm 0.0051$               & $126.87 \pm 0.3$  & $29.37\pm 0.29$    & $170.54 \pm 0.85$   \\
S62$^{\mathrm{b}}$   & $740.079 \pm 0.4125$  & $0.976 \pm 0.002$                 & $72.76 \pm 4.58$  & $42.62 \pm 0.4$    & $122.61 \pm 0.57$   \\
S4711$^{\mathrm{b}}$ & $619.208 \pm 12.3759$ & $0.768 \pm 0.030$                 & $114.71 \pm 2.92$ & $131.59 \pm 3.09$  & $20.10 \pm 3.72$    \\
S4713$^{\mathrm{b}}$ & $1653.42 \pm 78.1744$ & $0.351 \pm 0.059$                 & $111.07 \pm 1.66$ & $301.97 \pm 8.02$  & $195.06 \pm 5.15$   \\
S4714$^{\mathrm{b}}$ & $841.355 \pm 2.4752$  & $0.985 \pm 0.011$                 & $127.70 \pm 0.28$ & $357.25 \pm 0.80$  & $129.28 \pm 0.63$   \\
S4715$^{\mathrm{b}}$ & $1187.26 \pm 90.5503$ & $0.247 \pm 0.040$                 & $129.80 \pm 3.72$ & $359.99 \pm 5.38$  & $282.15 \pm 2.92$   \\
S4716$^{\mathrm{c}}$ & $400.154 \pm 4.1253$  & $0.756 \pm 0.02$                  & $161.13 \pm 2.80$ & $2.25 \pm 0.02$    & $153.55 \pm 1.54$   \\
\hline
\end{tabular}}
\\[2pt]
\begin{minipage}{0.8\textwidth}
\footnotesize
$^{\mathrm{a}}$ Updated data adopted from \cite{GRAVITY2024}.\\
$^{\mathrm{b}}$ Data taken from \cite{Peißker_2020b,Peißker_2020a}.\\
$^{\mathrm{c}}$ Orbital parameters adopted from \cite{Peißker_2022}.\\
$^{\mathrm{d}}$ Data adopted from \cite{Gillessen_2017}.
\end{minipage}
\end{center}

\section{Astrometric shift of S-stars in the sky caused by different dark matter profiles}
In this section, we estimate the astrometric shift of the orbits of the 13 S-stars in the sky in case of pure Schwarzschild as well as Schwarzschild plus dark matter profile.

\noindent
\begin{center}
\captionof{table}{Astrometric shift in the sky caused by the 13 S-stars considered in this work in Schwarzschild as well as Schwarzschild plus dark matter profiles.}
\label{tabC}
\resizebox{\textwidth}{!}{%
\begin{tabular}{ccccccccc}
\hline
Star &
  \begin{tabular}[c]{@{}c@{}}GR (Schwarzschild)\\ ($\mu$as)\end{tabular} &
  \begin{tabular}[c]{@{}c@{}}GR + Plummer\\ ($\mu$as)\end{tabular} &
  \begin{tabular}[c]{@{}c@{}}GR + GPC\\ ($\mu$as)\end{tabular} &
  \begin{tabular}[c]{@{}c@{}}GR + BWC\\ ($\mu$as)\end{tabular} &
  \begin{tabular}[c]{@{}c@{}}GR + GS\\ ($\mu$as)\end{tabular} &
  \begin{tabular}[c]{@{}c@{}}GR + SIDM ($r^2$)\\ ($\mu$as)\end{tabular} &
  \begin{tabular}[c]{@{}c@{}}GR + SIDM ($r^3$)\\ ($\mu$as)\end{tabular} &
  \begin{tabular}[c]{@{}c@{}}GR + FDM\\ ($\mu$as)\end{tabular} \\ \hline
S2    & $312.061$  & $294.346$ & $288.562$ & $287.024$  & $285.532$  & $294.438$  & $294.585$  & $297.477$  \\
S21   & $196.335$  & $40.8569$ & $41.990$ & $58.141$   & $102.475$  & $13.971$   & $3.902$    & $44.633$   \\
S23   & $103.454$  & $153.377$ & $134.661$ & $96.087$   & $9.243$    & $285.533$  & $262.032$  & $386.569$  \\
S29   & $909.084$  & $783.669$ & $723.041$ & $752.967$  & $802.715$  & $606.257$  & $680.743$  & $374.434$  \\
S38   & $56.657$   & $47.154$ & $45.731$ & $45.628$   & $46.647$   & $46.860$   & $46.926$   & $47.942$   \\
S55   & $71.297$   & $63.621$ & $60.956$ & $60.223$   & $60.076$   & $64.594$   & $64.71$    & $65.923$   \\
S60   & $157.744$  & $356.339$ & $485.068$ & $336.879$  & $77.582$   & $1069.254$ & $792.093$  & $2248.823$ \\
S62   & $1924.399$ & $1921.299$ & $1917.647$ & $1916.153$ & $1910.889$ & $1921.187$ & $1921.262$ & $1922.382$ \\
S4711 & $211.633$  & $205.852$ & $200.521$ & $198.609$  & $195.27$   & $206.992$  & $207.18$   & $208.175$  \\
S4713 & $101.696$  & $116.331$ & $81.694$ & $58.705$   & $3.31$     & $160.387$  & $160.577$  & $171.358$  \\
S4714 & $2539.257$ & $2536.367$ & $2533.101$ & $2531.898$ & $2527.314$ & $2535.857$ & $2535.909$ & $2537.315$ \\
S4715 & $78.184$   & $3.437$ & $10.333$  & $14.020$   & $31.353$   & $7.640$    & $7.326$    & $16.175$   \\
S4716 & $72.129$   & $71.717$ & $70.776$ & $70.364$   & $69.309$   & $71.820$   & $71.841$   & $71.909$   \\ \hline
\end{tabular}%
}
\end{center}

\section{Contour Plots}
The contour plots demonstrating the variation of ratio of precession induced by different dark matter distributions to that of Schwarzschild precession with semi-major axis and eccentricity are presented below.

\begin{figure}
    \centering
    \begin{subfigure}[b]{0.39\textwidth}
        \centering
        \includegraphics[width=\linewidth]{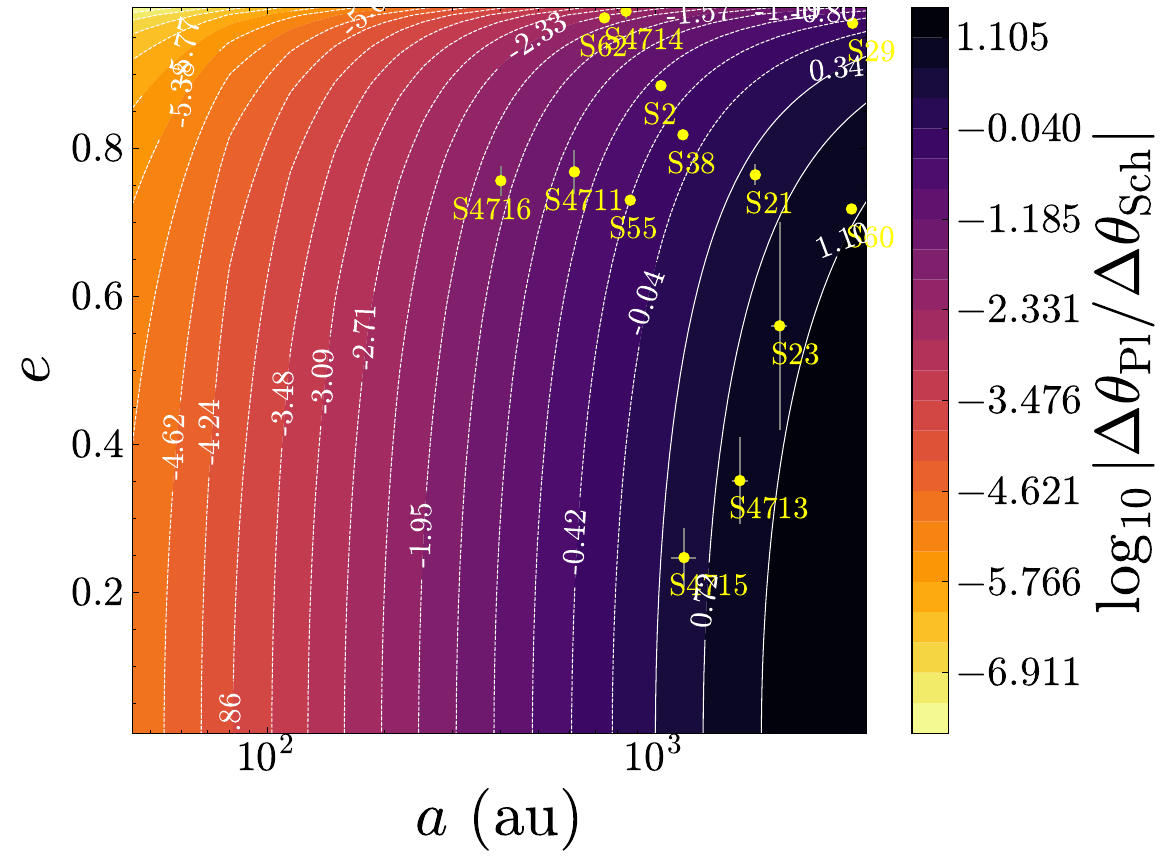}
        \caption{}
        \label{fig2}
    \end{subfigure}
    \begin{subfigure}[b]{0.39\textwidth}
        \centering
        \includegraphics[width=\linewidth]{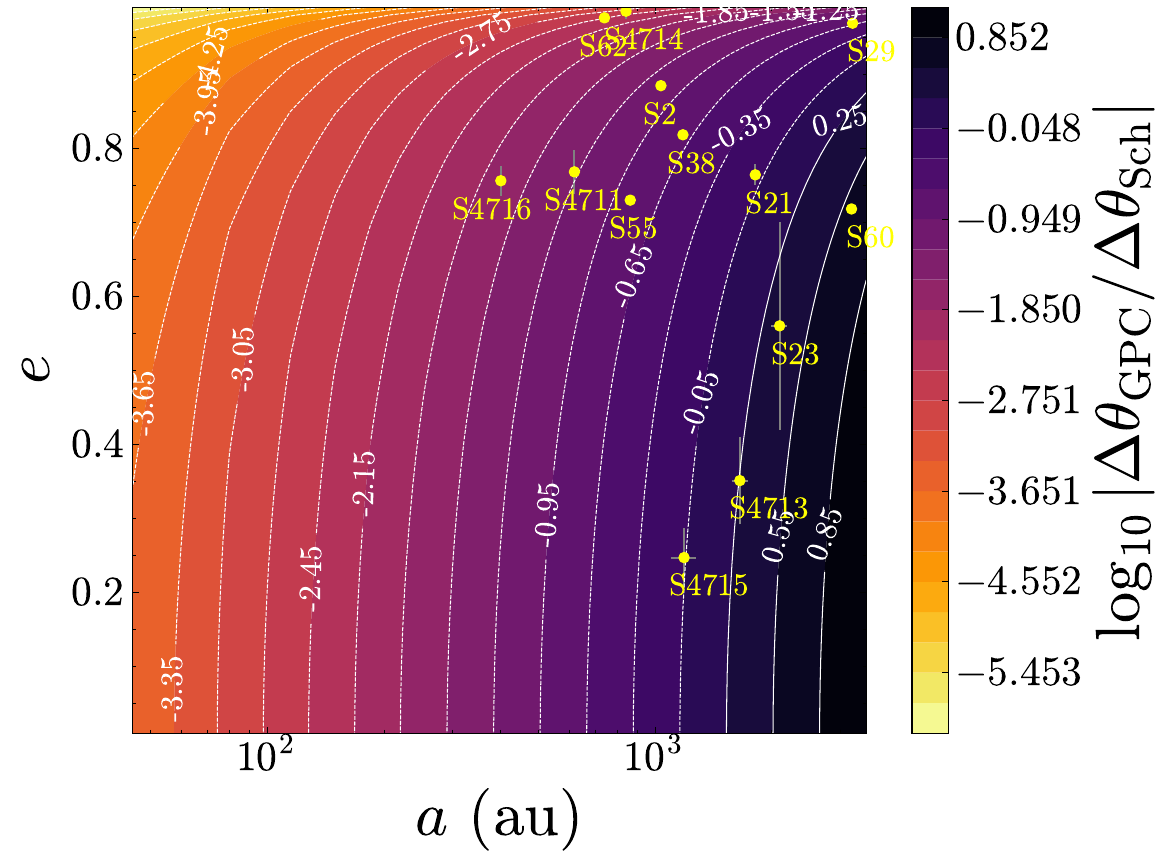}
        \caption{}
        \label{fig2a}
    \end{subfigure}\\
    \begin{subfigure}[b]{0.39\textwidth}
        \centering
        \includegraphics[width=\linewidth]{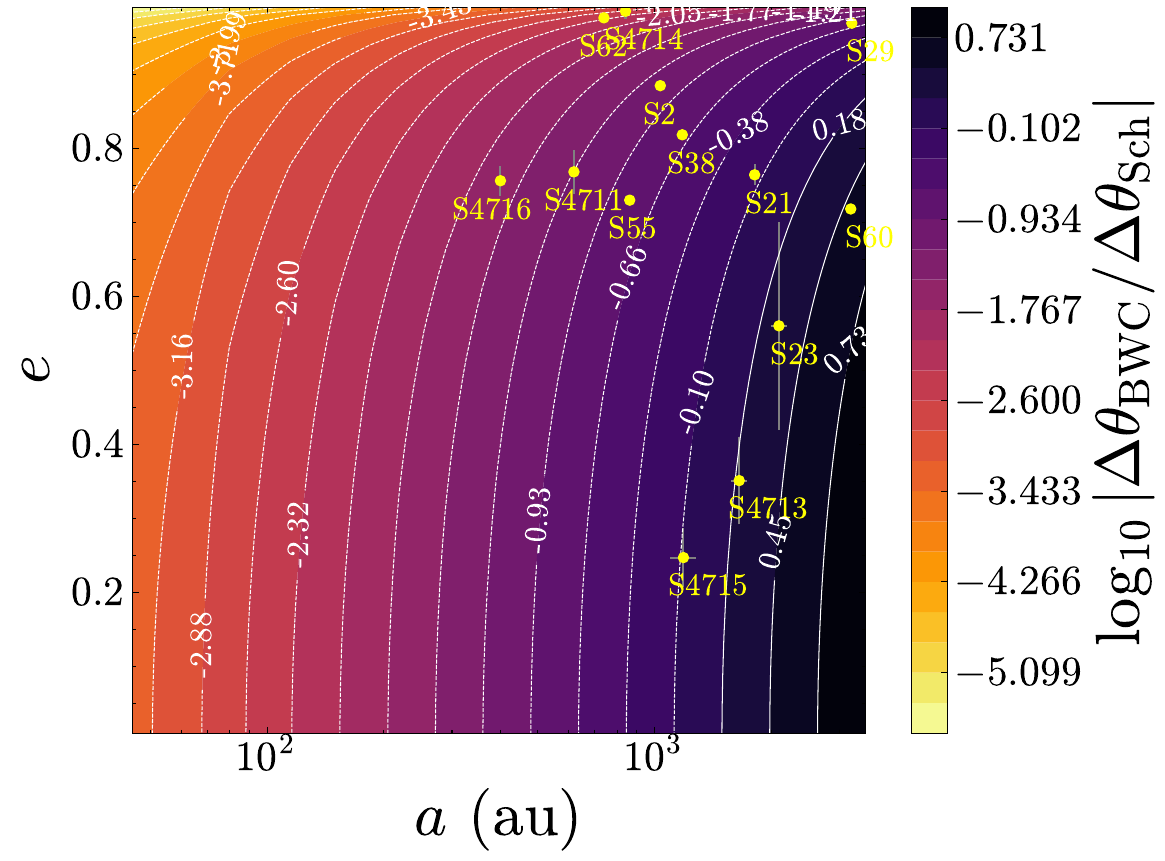}
        \caption{}
        \label{fig3}
    \end{subfigure}
    \begin{subfigure}[b]{0.39\textwidth}
        \centering
        \includegraphics[width=\linewidth]{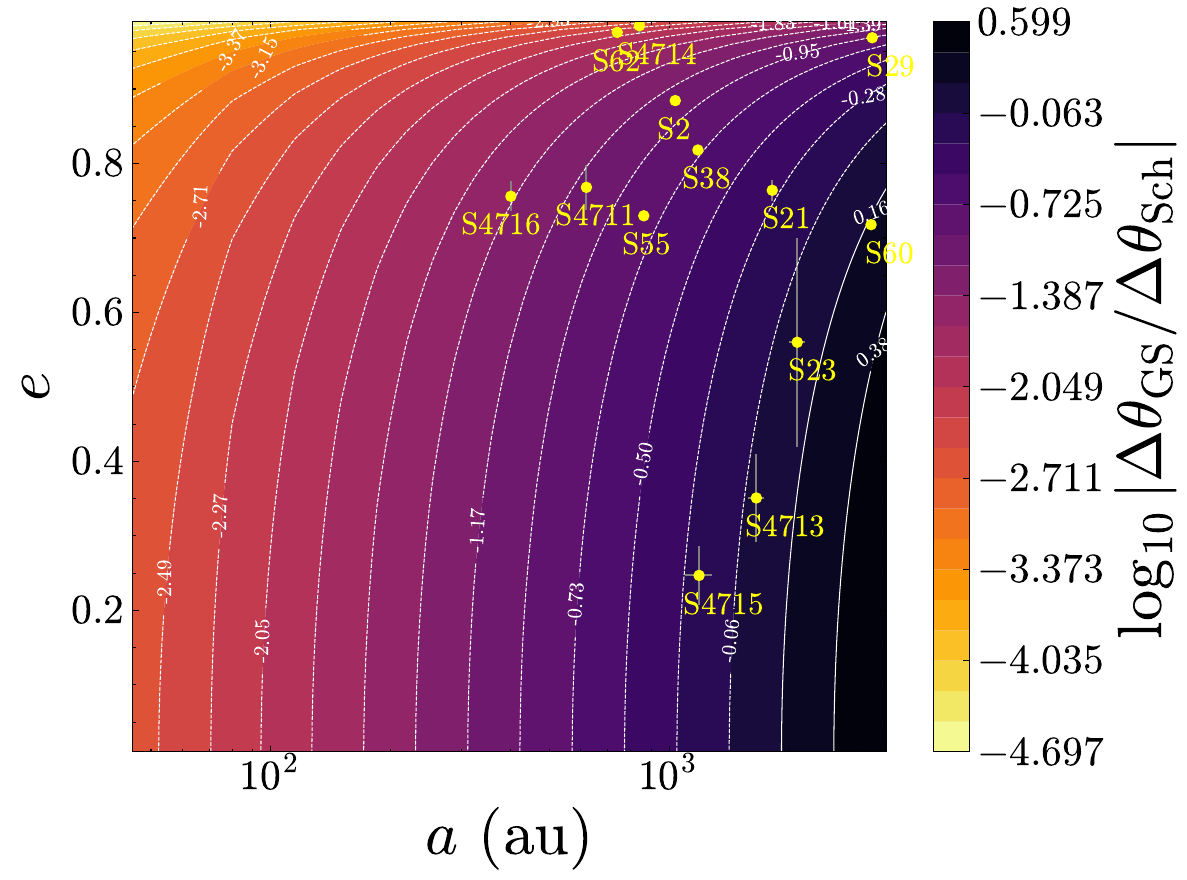}
        \caption{}
        \label{fig4}
    \end{subfigure}\\
    \begin{subfigure}[b]{0.39\textwidth}
        \centering
        \includegraphics[width=\linewidth]{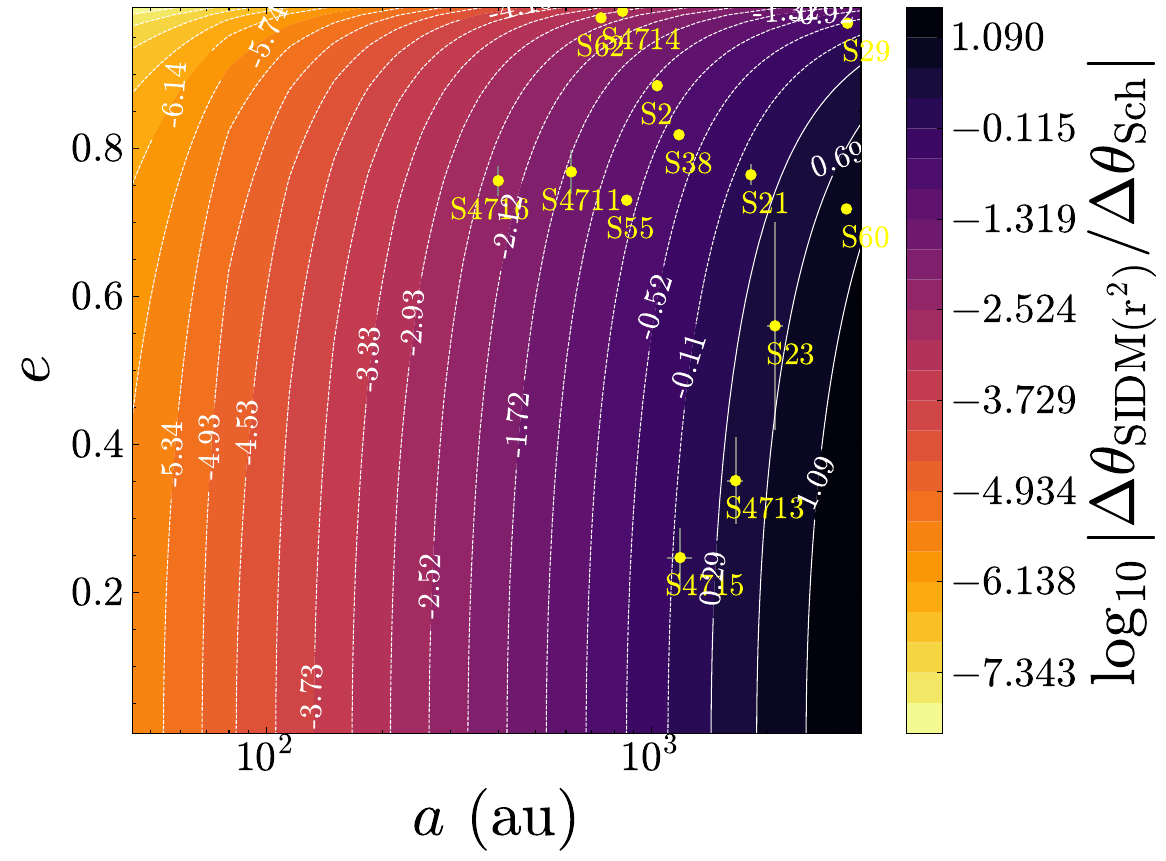}
        \caption{}
        \label{fig5}
    \end{subfigure}
    \begin{subfigure}[b]{0.39\textwidth}
        \centering
        \includegraphics[width=\linewidth]{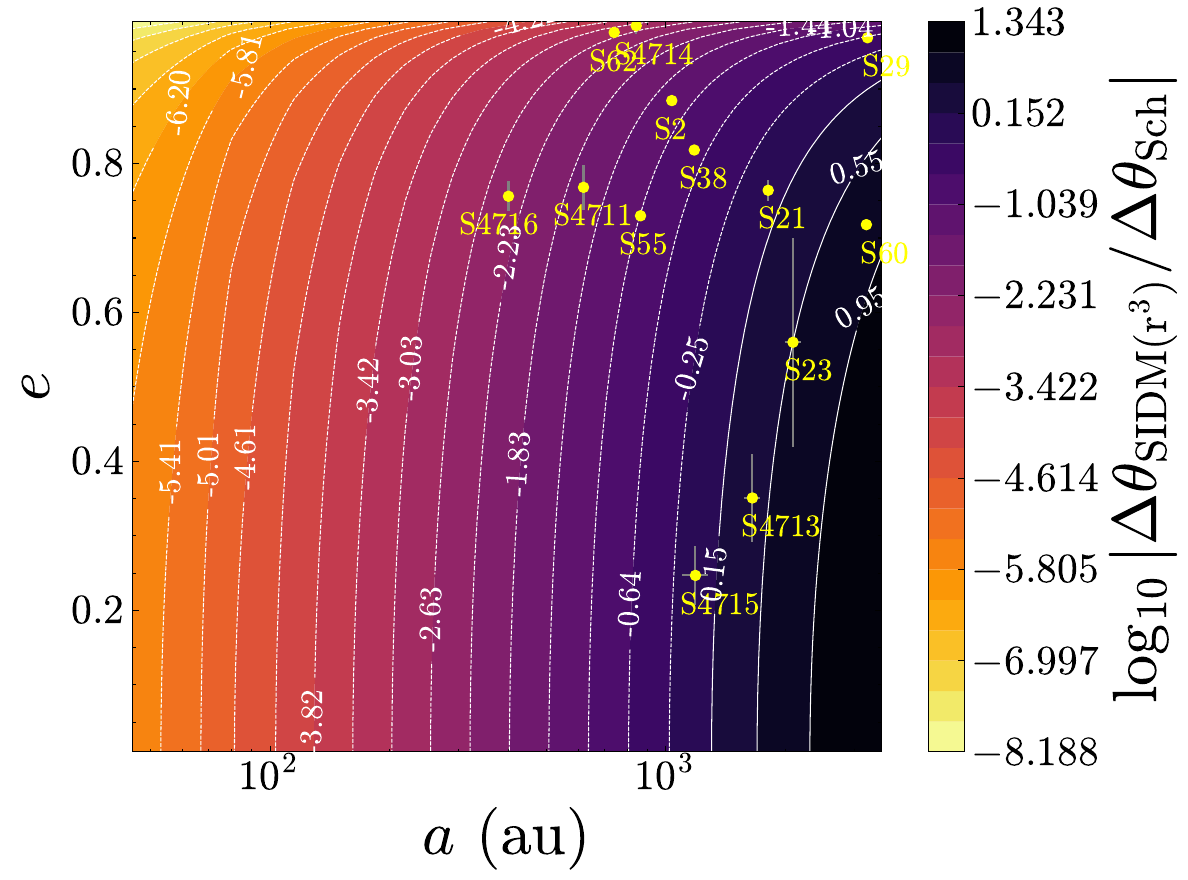}
        \caption{}
        \label{fig6}
    \end{subfigure}\\
    \begin{subfigure}[b]{0.39\textwidth}
        \centering
        \includegraphics[width=\linewidth]{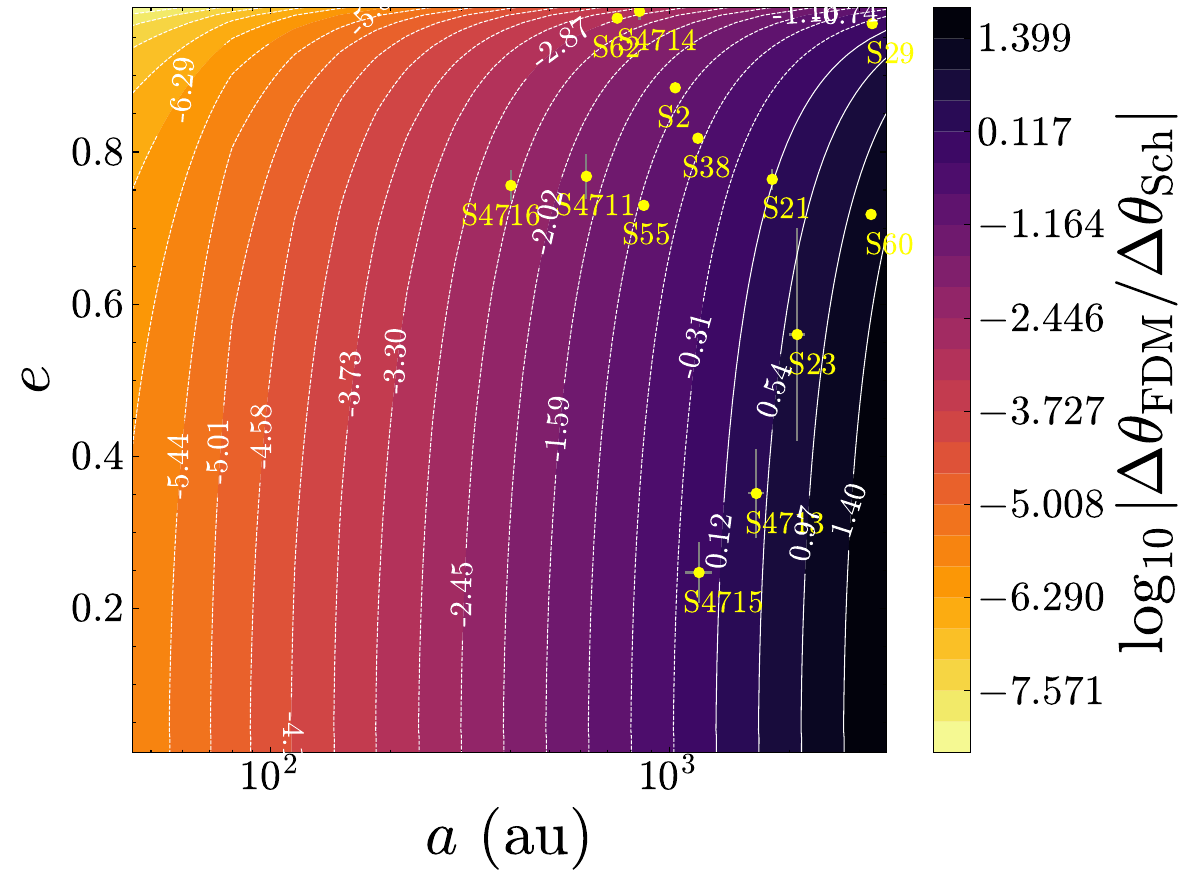}
        \caption{}
        \label{fig7}
    \end{subfigure}
    
\caption{The ratio of magnitude of precession per period in different profiles to that of Schwarzschild for $a$ values ranging from $45$ au to $3500$ au and $e$ ranging from $0.01$ to $0.99$. The locations of the S-stars (with $a$ up to $3500$ au) presented in Table \ref{tabA} are displayed with their uncertainties.}
\label{fig1}
\end{figure}


\bsp	
\label{lastpage}
\end{document}